\documentclass[sigconf,nonacm, pdfa]{acmart}

\usepackage[a-2b]{pdfx}
\usepackage{latexsym}
\usepackage{amsfonts}
\usepackage{amsmath}
\usepackage{amssymb}    
\usepackage{color}
\usepackage{epsfig}
\usepackage{xspace}
\usepackage{graphicx}
\usepackage{balance}  
\usepackage{balance}
\usepackage{epsfig}
\usepackage{multirow}
\usepackage{url}
\usepackage{makecell}
\usepackage{colortbl}
\usepackage{caption}
\usepackage{subcaption}
\usepackage{enumitem}
\usepackage{listings}

\newcommand\vldbdoi{10.14778/3705829.3705847}
\newcommand\vldbpages{308 - 321}
\newcommand\vldbvolume{18}
\newcommand\vldbissue{2}
\newcommand\vldbyear{2024}
\newcommand\vldbauthors{Xiaoke Zhu, Min Xie, Ting Deng, Qi Zhang}
\newcommand\vldbtitle{\shorttitle} 
\newcommand\vldbavailabilityurl{https://github.com/SICS-Fundamental-Research-Center/HyperBlocker}
\newcommand\vldbpagestyle{empty}

\usepackage[normalem]{ulem}
\setlist{topsep=0pt,noitemsep} \setitemize[1]{label=$\circ$}

\definecolor{dkgreen}{rgb}{0,0.6,0}
\definecolor{gray}{rgb}{0.5,0.5,0.5}
\definecolor{mauve}{rgb}{0.58,0,0.82}

\usepackage{algpseudocode}
\usepackage{newfloat}
\DeclareFloatingEnvironment[fileext=frm,placement={!ht},name=Frame]{myfloat}
\usepackage{listings}
\usepackage{epsfig}
\usepackage{multirow}
\usepackage{url}

\usepackage{enumitem}
\setlist{topsep=0pt,noitemsep} \setitemize[1]{label=$\circ$}

\newcommand{\eat}[1]{}

\newcommand{\sstab}{\rule{0pt}{8pt}\\[-1.8ex]}

\newcommand{\ra}{\rightarrow}

\newcommand{\bi}{\begin{itemize}}
\newcommand{\ei}{\end{itemize}}
        {\end{itemize}} 

\newcommand{\be}{\begin{enumerate}}
\newcommand{\ee}{\end{enumerate}}
\newcommand{\beqn}{\begin{eqnarray}}
\newcommand{\eeqn}{\end{eqnarray}}

\newcommand{\stitle}[1]{\vspace{1.6ex}\noindent{\bf #1}}
\newcommand{\etitle}[1]{\vspace{1ex}\noindent{\underline{\em #1}}}
\newcommand{\eetitle}[1]{\vspace{0.8ex}\noindent{{\em #1}}}

\newcommand{\ie}{\emph{i.e.,}\xspace}
\newcommand{\eg}{\emph{e.g.,}\xspace}
\newcommand{\wrt}{\emph{w.r.t.}\xspace}

\newcommand{\kwlog}{\emph{w.l.o.g.}\xspace}

\newcommand{\reftab}[1]{Table \ref{#1}}

\newenvironment{proofS}{
        \vspace{1ex}
        {\noindent\bf Proof sketch:\ }}{\eop\vspace{1ex}}

\newcommand{\kw}[1]{{\ensuremath {\mathsf{#1}}}\xspace}

\newcounter{ccc}

\newcommand{\MD}{\kw{MD}}
\newcommand{\MDs}{\kw{MDs}}

\newcommand{\R}{{\mathcal R}}
\newcommand{\B}{{\mathcal B}}
\newcommand{\A}{{\mathcal A}}

\newcommand{\Q}{{\mathcal Q}}

\newcommand{\E}{{\mathcal E}}

\newcommand{\eop}{\hspace*{\fill}\mbox{$\Box$}}     

\newcounter{example}
\renewcommand{\theexample}{\arabic{example}}
\newenvironment{example}{
        \vspace{1.5ex}
        \refstepcounter{example}
        {\noindent\bf Example \theexample:}}{
        \eop\vspace{1.5ex}}

\newcommand{\nthesection}{\arabic{section}}

\newcounter{theorem}
\renewcommand{\thetheorem}{\arabic{theorem}}
\newcounter{prop}
\renewcommand{\theprop}{\arabic{theorem}}
\newcounter{lemma}
\renewcommand{\thelemma}{\arabic{theorem}}
\newcounter{cor}
\renewcommand{\thecor}{\arabic{theorem}}

\newenvironment{prop}{\begin{em}
        \refstepcounter{theorem}
        {\vspace{1.5ex}\noindent \bf Proposition \theprop:}}{
        \end{em}\eop\vspace{1.5ex}}
\newcounter{definition}[section]
\renewcommand{\thedefinition}{\nthesection.\arabic{definition}}

\newcounter{alg}[section]
\renewcommand{\thealg}{\nthesection.\arabic{alg}}

\newcounter{arule}
\renewcommand{\thearule}{\arabic{arule}}

\newcounter{claim}
\renewcommand{\theclaim}{\arabic{claim}}

\newcommand{\tbf}{\textbf{\textcolor{red}{X}}\xspace}

\newcommand{\REEs}{\kw{REEs}}

\renewcommand{\texttt}[1]{{\small\textsf{#1}}}

\definecolor{gray}{rgb}{0.5,0.5,0.5}

\newcommand{\T}{{\mathcal T}}

\newcommand{\N}{{\mathcal N}}

\newcommand{\D}{{\mathcal D}}

\renewcommand{\P}{{\mathcal P}}

\newcommand{\revise}[1]{#1}
\newcommand{\reviseX}[1]{#1}
\newcommand{\reviseS}[1]{{\color{black}{#1}}}
\newcommand{\reviseC}[1]{#1}
\newcommand{\reviseY}[1]{#1}
\newcommand{\improve}[1]{{\color{cyan}{#1}}}

\newcommand{\CR}[1]{#1}

\newcommand{\warn}[1]{\textcolor{black}{#1}}

\makeatletter
\newcommand\figcaption{\def\@captype{figure}\caption}
\newcommand\tabcaption{\def\@captype{table}\caption}
\makeatother

\newcommand{\eid}{\kw{eid}}

\usepackage[ruled,vlined]{algorithm2e}

\newcommand{\tabincell}[2]{\begin{tabular}
		{@{}#1@{}}#2\end{tabular}}

\newcommand{\Hyper}{\kw{HyperBlocker}}

\newcommand{\HP}{\kw{HP}}
\newcommand{\HR}{\kw{HR}}

\newlist{myitemize}{itemize}{3}
\setlist[myitemize,1]{label=$\circ$,leftmargin=2.6ex}
\setlist[myitemize,2]{label=$\bullet$,leftmargin=3.8ex}
\setlist[myitemize,3]{label=$\diamond$, leftmargin=3.2ex}
\newenvironment{mbi}{\begin{myitemize}
        \setlength{\topsep}{0.8ex}\setlength{\itemsep}{0.36ex}\vspace{0.6ex}}
        {\end{myitemize}\vspace{0.8ex}}

\newcommand{\mei}{\end{myitemize}\vspace{0.6ex}}

\title{HyperBlocker: Accelerating Rule-based Blocking in Entity Resolution using GPUs}

\author{Xiaoke Zhu}
\affiliation{
	\institution{Beihang University \country{China}}
}
\affiliation{
	\institution{Shenzhen Institute of \\Computing Sciences 
		\country{China}}
}
\email{zhuxk@buaa.edu.cn}
\author{Min Xie}
\affiliation{
	\institution{Shenzhen Institute of \\Computing Sciences \country{China}}
}
\email{xiemin@sics.ac.cn}
\authornote{Corresponding author}
\author{Ting Deng}
\affiliation{
	\institution{Beihang University \country{China}}
}
\email{dengting@act.buaa.edu.cn}
\author{Qi Zhang}
\affiliation{
	\institution{Meta Platforms \country{USA}}
}
\email{qizhang@meta.com}

\begin{document}

        \fancyhead{}

	\begin{abstract}
		This paper studies rule-based blocking in Entity Resolution (ER).
		We propose \Hyper, a  GPU-accelerated system for blocking in ER.
		As opposed to previous blocking algorithms and parallel blocking solvers,
		\Hyper employs a pipelined architecture to overlap data transfer and GPU operations.
		It generates a data-aware and rule-aware execution plan on CPUs, for specifying how rules are evaluated, and
		develops a number of hardware-aware optimizations to achieve massive parallelism on GPUs.
		
  Using real-life datasets, we show that \Hyper is at least \reviseY{6.8}$\times$ and  \reviseY{9.1}$\times$ faster than prior CPU-powered distributed systems
		and GPU-based ER solvers, respectively.
		Better still, by combining  \Hyper with the state-of-the-art ER matcher,
		we can speed up the overall ER process by at least 30\%  with comparable accuracy. 
	\end{abstract}

	\maketitle

	\setcounter{page}{1}
    \pagestyle{plain}

        \pagestyle{\vldbpagestyle}
        \begingroup\small\noindent\raggedright\textbf{PVLDB Reference Format:}\\
        \vldbauthors. \vldbtitle. PVLDB, \vldbvolume(\vldbissue): \vldbpages, \vldbyear.\\
        \href{https://doi.org/\vldbdoi}{doi:\vldbdoi}
        \endgroup
        \begingroup
        \renewcommand\thefootnote{}\footnote{\noindent
        This work is licensed under the Creative Commons BY-NC-ND 4.0 International License. Visit \url{https://creativecommons.org/licenses/by-nc-nd/4.0/} to view a copy of this license. For any use beyond those covered by this license, obtain permission by emailing \href{mailto:info@vldb.org}{info@vldb.org}. Copyright is held by the owner/author(s). Publication rights licensed to the VLDB Endowment. \\
        \raggedright Proceedings of the VLDB Endowment, Vol. \vldbvolume, No. \vldbissue\ %
        ISSN 2150-8097. \\
        \href{https://doi.org/\vldbdoi}{doi:\vldbdoi} \\
        }\addtocounter{footnote}{-1}\endgroup
        
        \ifdefempty{\vldbavailabilityurl}{}{
        \vspace{.3cm}
        \begingroup\small\noindent\raggedright\textbf{PVLDB Artifact Availability:}\\
        The source code, data, and/or other artifacts have been made available at \url{\vldbavailabilityurl}.
        \endgroup
        }

	\setcounter{page}{1}

\section{Introduction}
\label{sec-intro}

Entity resolution (ER), 
also known as record linkage, data deduplication, merge/purge
and record matching, 
is to identify tuples that refer to the same real-world entity. 
It is a routine operation in many data cleaning and integration tasks, 
 such as detecting duplicate commodities~\cite{e-commerce-er} and finding duplicate customers~\cite{deng2022deep}.

Recently, with the rising popularity of deep learning (DL) models, 
research efforts have been made to apply DL techniques to ER. Although these DL-based approaches have shown impressive accuracy, they also come with high training/inference costs, due to the large number of parameters.
\eat{\eg 
in our preliminary experiment, 
the state-of-the-art ER model, \kw{Ditto}~\cite{ditto}, takes more than 2000 seconds to process 2 million pairs of tuples. 
it was reported in~\cite{}\tbf that, the state-of-the-art ER model, \kw{Ditto}, takes more than \tbf seconds to process a dataset with only \tbf tuples. 
}
Despite the effort to reduce parameters,
the growth in the size of DL models is still an inevitable trend, 
leading to the increasing time for making matching decisions.

In the worst case, ER solutions have to spend quadratic time examining all pairs of tuples.
As reported by Thomson Reuters,
an ER project can take 3-6 months, 
mainly due to the scale of data~\cite{dis-dup}.
To accelerate, most ER solutions divide ER into two phases:
(a) a blocking phase, where a blocker discards unqualified pairs that are guaranteed to refer to distinct entities, 
and (b) a matching phase, where a matcher compares the remaining pairs to finally decide whether they are \emph{matched}, \ie 
refer to the same entity.
The blocking phase is particularly useful when dealing with large data
and 
\reviseX{``is the crucial part of ER with
respect to time efficiency and scalability''}~\cite{blocking-filtering-survey}.
\looseness=-1

To cope with the volume of big data,
considerable research has been conducted on blocking techniques.
As surveyed in~\cite{li2020survey, blocking-filtering-survey},
we can divide blocking methods into \emph{rule-based} \cite{dedoop,dis-dup,rules-blocking,papadakis2011compare,gu2004adaptive} or \emph{DL-based} \cite{DeepER,DeepBlocking,zhang2020autoblock,javdani2019deepblock},
both have their strengths and limitations.

\begin{figure}[t]
\vspace{-1.2ex}
    \centering
	\subcaptionbox{
        Execution time
	}{
		\includegraphics[width=0.23\textwidth]{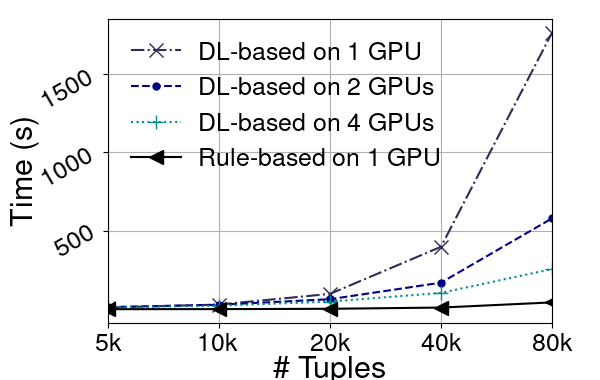} \vspace{-1.5ex}
	} 
	\hspace{-4mm} 
	\subcaptionbox{
		Memory cost
	}{
        \includegraphics[width=0.23\textwidth]{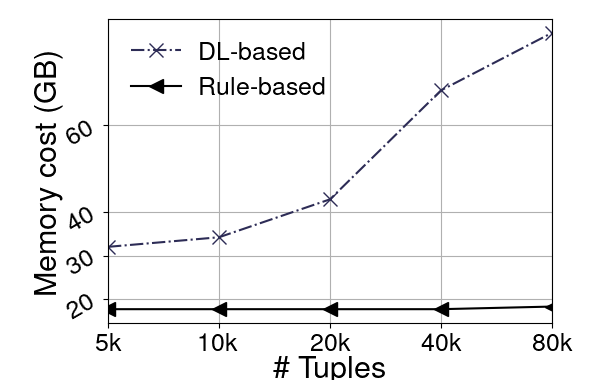} \vspace{-1.5ex}
	}	
    \vspace{-3ex}
    \caption{\reviseX{DL-based blocking vs. rule-based blocking}} \label{exp:motivation}
    \vspace{-4ex}
\end{figure}

DL-based blocking methods typically utilize pre-trained DL models to generate embeddings for tuples and discard tuple pairs with low similarity scores. 
While DL-based blocking can enhance ER by parallelizing computation and leveraging GPU acceleration~\cite{Faiss}, it often comes with long runtime and high memory costs. 
To justify this, we conducted a detailed analysis on \kw{DeepBlocker}~\cite{DeepBlocking},
the state-of-the-art \reviseY{(SOTA)} DL-based blocker in Figure~\ref{exp:motivation}.
We picked a rule-based blocker (a prototype of our method) with comparable accuracy with \kw{DeepBlocker}
and compared \reviseC{their} runtime and memory. The evaluation was conducted on a machine equipped with V100 GPUs using the {\tt Songs} dataset~\cite{DeepMatcher}, varying the number of tuples. \eat{We employ one GPU for the rule-based blocker while employing one or four GPU(s) for \kw{DeepBlocker}.}
\reviseX{When running on one GPU,}
the runtime of \kw{DeepBlocker} increases substantially when the number of tuples exceeds 40k.
Worse still, it consumes excessive memory due to the large embeddings and intermediate results during similarity computation. 
\reviseX{
Although the runtime of \kw{DeepBlocker} can be reduced by using more GPUs,
the issue remains, \eg 
even with four GPUs in Figure~\ref{exp:motivation}(a),
\kw{DeepBlocker} is still slower than the rule-based blocker that runs on one GPU}.
\looseness=-1

\revise{
In contrast, rule-based blocking methods demonstrate potential for achieving scalability by leveraging multiple \emph{blocking rules}. Each rule employs various comparisons with logical operators such as AND, OR, and NOT to discard unqualified tuple pairs. 
For instance, a blocking rule for books may state ``If titles match and the number of pages match, then the two books match''~\cite{konda2016magellan}. We refer to the comparisons in this rule as \emph{equality comparisons}, as they require exact equality. Another example, referred to as \emph{similarity comparisons}, is presented in \cite{papadakis2011compare}, which adopts the Jaccard similarity to determine whether a pair of tuples requires further matching. Rule-based approaches complement DL-based approaches by providing flexibility, explainability, and scalability in the blocking process~\cite{ShallowBlocker}. Moreover, by incorporating domain knowledge into blocking rules, these approaches can readily adapt to different domains.
}
\looseness=-1


\eat{
For instance, let's consider a relation displayed in Table\ref{tab:products} that represents products, having attributes like pname, price, owner, desc, and ip. Using rule-based methods, the company can identify that the products \tbf and \tbf potentially refer to the same product due to their matching \tbf and similar \tbf. In contrast, in DL-based methods, each tuple in the \kw{Products} table is converted into an embedding vector. Subsequently, a similarity search model, such as Faiss\cite{Faiss}, is employed to search the embedded \kw{Products} and identify pairs of vectors that exhibit similarity. The original tuples corresponding to these pairs are then outputted as the set of candidate matches.
}

\begin{example}\label{example:1}
As a critical step for data consistency, 
an e-commerce company (\eg Amazon~\cite{amazon}) conducts ER for products, to enhance operations for \revise{\eg product listings and inventory management.} \looseness=-1

To identify duplicate products,
the blocking rule $\varphi_1$ may fit.

\vspace{0.36ex}
\noindent
$\varphi_1$: Two products are potentially matched if (a) they have same color and price,
(b) they are sold at same store, 
(c) their names are similar. \looseness=-1

\vspace{0.36ex}
Here $\varphi_1$ is a conjunction of attribute-wise comparisons,
where both equality (parts (a) and (b))
and similarity comparisons  (part (c))
are involved.
In Section~\ref{sec-background},
we will formally define $\varphi_1$.
\end{example}
\looseness=-1

\begin{figure}[t]
	\centering
	\includegraphics[width=1\linewidth]{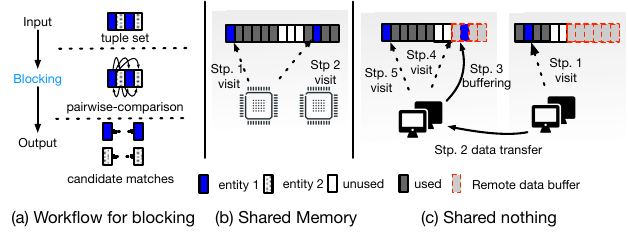}
        \vspace{-6.8ex}
	\caption{\reviseS{Shared memory vs shared nothing architectures}}
	\label{fig:shared_memory_vs_shared_nothing}
\vspace{-4ex}
\end{figure} 

\eat{Considering that DL excels in effectiveness and rule excels in efficiency, it is reasonable to use rule-based methods in the blocking phase and DL-based methods in the matching phase.
Motivated by this, }
Rule-based blocking in ER has attracted a lot of attention (surveyed in \cite{li2020survey,blocking-filtering-survey}).
However, most existing rule-based blockers are designed for CPU-based (shared nothing) architectures, leading to unsatisfactory performance. 
\reviseS{Typically, 
a blocker conducts pairwise comparisons on all pairs of tuples to obtain candidate matches (Figure~\ref{fig:shared_memory_vs_shared_nothing}(a)). 
In a shared-nothing architecture, 
data is partitioned and spread across a set of processing units.
Each unit independently blocks data using its local memory, which may lead to skewed partitions/computations and rising communication costs,
\eg in Figure~\ref{fig:shared_memory_vs_shared_nothing}(c),
the first unit is assigned more tuples than the second one and worse still,
two tuples that both refer to entity 1 are distributed to different partitions.
To avoid missing this match,
the second unit has to visit its local memory (Stp. 1) and transfer its data to the first unit (Stp. 2).
Then the first unit buffers the data (Stp. 3) and finally, conducts comparison locally (Stp. 4 \& 5). 
The shared memory architecture (Figure~\ref{fig:shared_memory_vs_shared_nothing}(b)) is the opposite: all data is accessible from all processing units,
allowing for efficient data sharing, collaboration between processing units and dynamic workload scheduling,
\eg the two tuples referred to entity 1 can be directly accessed by both units in Figure~\ref{fig:shared_memory_vs_shared_nothing}(b) (Stp. 1 \& 2).}
\eat{Consider that blocking is a task might involve fine-grained parallelism, frequent communication and is prone to skewness~\cite{blocking-filtering-survey}. 
The shared-memory naturally becomes the preferred choice.}
GPUs are typically based on shared memory architectures, offering promising opportunities to achieve blocking parallelism.
\eat{
Considering that blocking operations are computationally intensive by nature, we argue that GPUs based on shared memory architecture, providing massive parallelism, offer new avenues to improve the efficiency of rule-based blocking.}
\eat{
However, unlike DL-based blocking approaches, few rule-based methods leverage the massive parallelism offered by GPUs, despite their inherently low computational dependencies and greater potential.
}
\revise{However, 
unlike DL-based approaches, few rule-based methods support the massive parallelism offered by GPUs, despite
their greater potential in parallel scalability~(see Figure~\ref{exp:motivation}).}
\looseness=-1

\vspace{0.6ex}
To make practical use of rule-based blocking, several questions have to be
answered.
\reviseY{Can we parallelize it under a share memory architecture, utilizing massive parallelism of a GPU?
Can we explore characteristics of GPUs and CPUs, to effectively collaborate them? }

\eat{
Can we parallelize the existing rules-based methods on GPU or across multiple processors of GPUs?
Can we make the most resources of a GPU 
to exploit GPU hardware characteristics for these methods?
}

\vspace{-0.8ex}
\stitle{HyperBlocker}.
To answer these, we develop \Hyper, \revise{a GPU-accelerated system for 
 rule-based blocking in Entity Resolution.}
As proof of concept,
we adopt matching dependencies (\MDs)~\cite{fan2011dynamic} for rule-based blocking.
As a class of rules developed for record matching, \MDs are defined as a conjunction of  (similarity) predicates and support both equality and similarity comparisons.
Compared with prior works,
\Hyper has the following unique features. \looseness=-1

\etitle{(1) A pipelined architecture}.
\kw{HyperBlocker} adopts \revise{an architecture}
that pipelines the memory access from/to CPUs for data transfer, and 
operations on GPUs for rule-based blocking. 
In this way,
the data transfer and the computation on GPUs can be overlapped.
\looseness=-1

\etitle{(2) Execution plan on CPUs}.
To effectively filter unqualified pairs, 
\revise{blocking must be optimized for the underlying data (resp. blocking rules) \reviseS{for both equality and similarity comparisons};
in this case, we say that the blocking is \emph{data-aware} (resp. \emph{rule-aware}}).
To our knowledge,
\reviseS{prior methods either fail to consider data/rule-awareness 
or cannot handle arbitrary comparisons well.} 
\Hyper designs an execution plan generator to 
warrant efficient rule-based blocking. \looseness=-1

\etitle{(3) Hardware-aware parallelism on GPUs}.
Due to different characteristics of CPUs and GPUs,
a naive approach that applies existing CPU-based blocking on GPUs makes
substantial processing capacity untapped. 
We develop a variety of GPU-based parallelism strategies, designated for rule-based blocking,
by exploiting the
hardware characteristics of GPUs,
to achieve massive parallelism.

\etitle{(4) Multi-GPUs collaboration}.
It is already hard to offload tasks on CPUs. 
\reviseX{This problem is even exacerbated under multi-GPUs, 
due to the complexities of task decomposition, (inter-GPU) resource management, and workload balancing.}
\kw{HyperBlocker} provides effective partitioning and scheduling strategies to scale with multiple GPUs. 

\eat{ 
\vspace{0.6ex}
With these attractive features,
\Hyper is capable of conducting rule-based blocking 
with shorter time and less memory,
as shown in Figure~\ref{exp:motivation}; more sophisticated experiments are shown later.}
\looseness=-1

\vspace{-0.3ex}
\stitle{Contribution \& organization.}
After reviewing background  in Section~\ref{sec-background},
we present \Hyper as follows:
(1)  its unique architecture and system overview (Section~\ref{sec-overview});
(2)  the rule/data-aware execution plan generator (Section \ref{sec-plan});
(3) the hardware-aware parallelism and 
the task scheduling strategy across GPUs~(Section \ref{sec-exec-model}); and
(4) an experimental study~(Section~\ref{sec-expt}).
\reviseX{Section~\ref{sec:related-work} presents  related work.}
\looseness=-1

Using real-life datasets, we
find the following: 
(a) \Hyper speedups prior distributed blocking systems and GPU baselines by at least \reviseY{6.8$\times$ and  9.1$\times$}, respectively.
(b) Combining \Hyper with the SOTA ER matcher saves at least 30\% time with comprable accuracy.
(c) \Hyper is scalable, \eg it can process 36M tuples in 1604s.
\revise{
(d) While promising, DL-based blocking methods are not always the best.
By carefully optimizing rule-based blocking on GPUs, we share valuable lessons/insights about when rule-based approaches can beat the DL-based ones and vice versa.}

\eat{
\begin{itemize}
    \item its unique architecture and system overview (Section~\ref{sec-overview});
    \item the rule/data-aware execution plan generator (Section \ref{sec-plan}).
    \item the hardware-aware parallelism optimizations and 
    the task scheduling strategy across multiple GPUs~(Section \ref{sec-exec-model}).
    \item an experimental study~(Section~\ref{sec-expt}). Using real-life and synthetic datasets, we
    find:
(a) \Hyper speedups prior distributed ER blocking systems and GPU baselines by at least 6.1$\times$ and  11.2$\times$, respectively.
(b) Combing \Hyper with the state-of-the-art ER matcher saves 26.5$\times$ time with comparable accuracy.
(c) \Hyper is scalable, \revise{\eg it can process 36M tuples in 1604s.} 
\end{itemize}
}

\begin{figure*}
\tabcaption{A relation $D$ of schema \kw{Products}, where the dash (``-'') denotes a missing value.}
\label{tab:products}
\begin{minipage}{18cm}
\begin{footnotesize}
		\begin{center}
        \scriptsize
			\begin{tabular}{c|c|c|c|c|c|c|c|}
				\cline{2-8}
				\kw{eid} & \kw{pno} & \kw{pname} & \kw{price} & \kw{sname} & \kw{description} & \kw{color} & \kw{saddress}       \\ 
                    \cline{2-8}
				$e_{1}$	& $t_1$ &  \tabincell{c}{Apple Mac Air}&  \$909    &  \tabincell{c}{Comp.  World} & Apple MacBook Air (13-inch, 8GB RAM, 256GB SSD) &  Gray  & \tabincell{c}{9 Barton Grove, McCulloughmouth} \\ 
                \cline{2-8} 
				$e_{2}$	& $t_2$  &ThinkPad &  -         & \tabincell{c}{Smith's Tech} & \tabincell{c}{ThinkPad E15, 15.6-inch full HD IPS display, Intel  Core  i5-1235U processor,  (16GB) RAM | 512GB PCIe SSD)} & Gray  & \tabincell{c}{Seg Plaza, Hua qiang North Road} \\ 
                \cline{2-8} 
				$e_{2}$	& $t_3$  &ThinkPad &  \$849         & \tabincell{c}{Smith's Tech} & \tabincell{c}{Lenovo  E15 Business ThinkPad, 15.6-inch full HD  IPS  display, 12 generation Intel Core i5, 16GB RAM, 512GB SSD} & Gray  & \tabincell{c}{Seg Plaza, Hua  qiang North Road} \\ 
                \cline{2-8} 
				$e_{1}$	& $t_4$  & MacBook Air  &  \$909        &  \tabincell{c}{Comp.  World} & \tabincell{c}{Apple 2022 MacBook  Air M2 chip 13-inch,8 GB RAM,256 GB SSD storage  gray} &  Gray  &  \tabincell{c}{-} \\ 
                \cline{2-8} 
                $e_{1}$	& $t_5$ &  MacBook Air&  \$909    &  \tabincell{c}{Comp.  World} & \tabincell{c}{-} & Gray  &  \tabincell{c}{Barton Grove, McCulloughmouth}\\                 
                \cline{2-8}         
			\end{tabular}
\end{center}
\end{footnotesize}
\end{minipage}
\vspace{-2ex}
\end{figure*}

\section{Preliminaries}
\label{sec-background}
We first review \CR{the \reviseX{notations}} for ER, blocking, and the GPU.

\vspace{-0.3ex}
\stitle{Relations.}
Consider a schema $R =
(\eid, A_1, \dotsc, \allowbreak A_n)$, where $A_i$
is an attribute ($i\in[1,n]$), and \eid is an entity id, such that
each tuple of $R$ represents an entity.
A relation $D$ of $R$ is a set of tuples of schema $R$. 
\looseness=-1

\vspace{-0.3ex}
\stitle{Entity resolution (ER).}
Given a relation $D$, ER is to identify all tuple pairs in $D$ that refer to the same real-life entity.
It returns a set of tuple pairs $(t_1, t_2)$ of $D$
that are identified
as {\em matches}. 
If $t_1$ does not match $t_2$, $(t_1, t_2)$
  is referred to as a \emph{mismatch}.

\vspace{0.6ex}
Most existing methods typically conduct ER in three steps:

\etitle{(1) Data partitioning.}
The tuples in  relation $D$ are divided into multiple data partitions,
namely $P_1,P_2,..., P_m$,
so that tuples of similar entities tend to be put into the same data partition.

\etitle{(2) Blocking.}
Each tuple pair $(t_1, t_2) $ \reviseC{from a partition $P$} is a potential match
that requires further verification.
To reduce cost,
a blocking method $\A_\kw{block}$ (\ie blocker) is often adopted to filter out those pairs 
that are definitely mismatches \emph{efficiently}, 
instead of directly verifying every tuple pair.
Denote the set of remaining pairs obtained from  $P$ by $\kw{Ca}(P) = \{(t_1, t_2) \in P \times P \mid (t_1, t_2) \text{ is not filtered by } \A_\kw{block}\}$.

\etitle{(3) Matching.}
\reviseC{For each pair in $\kw{Ca}(P)$},
an accurate (but expensive) matcher is applied,  to make final decisions of matches/mismatches.
\looseness=-1

\eat{
A candidate selection step often been split into two different
but related phrase: data partitioning and blocking. 
Data partitioning attempt to 
clusters similar entities into a
block collection $\B=\{B_1,B_2,...\}$ such that $\cup_{P\in\B}{P} = \E$.
Given an entity collection $\E$, a  function $f:\E\times\E\to\mathbb{B}$, 
blocking identifies all pairs of entity profiles in $\E$ 
that have $f$ return ture, \ie $\kw{Ca}(\E) = \{ (t,s)\in \E\Join\E, f(t,s) = true, t \neq s \}$.
}

\revise{
\vspace{-0.7ex}
\stitle{Our scope: blocking.}
Note that in some works, both steps (1) and (2) are called blocking.
To avoid ambiguity,
we follow~\cite{DeepBlocking}
and distinguish partitioning from blocking.
We mainly focus on \emph{blocking}, \ie \looseness=-1


\mbi
\item {\em Input}: A relation $D$ of the tuples of schema $R$, where the tuples in $D$ are divided into $m$ partitions $P_1, \ldots, P_m$.
\item {\em Output}: The set $\kw{Ca}(P_i)$ of candidate tuple pairs on each $P_i$.
\mei}

Although our work can be applied on data partitions generated by \emph{any} existing method,
we optimize over multiple data partitions,
by exploiting designated GPU acceleration techniques (Section~\ref{subsec:GPUs}).  \looseness=-1

\revise{
While blocking focuses more on efficiency and  matching focuses more on accuracy,
they can be used without each other,
\eg one can directly employ rules~\cite{fan2011dynamic} for ER or
 apply an ER matcher~\cite{ditto} on the Cartesian product of the entire partition.
 When blocking is used alone on a given partition $P$,
all tuple pairs in $\kw{Ca}(P)$ are identified as matches.
In Section~\ref{sec-expt},
we will test \Hyper with or without a matcher,
to elaborate the trade-off between efficiency and accuracy.
} \looseness=-1

\eat{
\stitle{Our problem setting.}
Much of the literature uses the terms "data partitioning" and "blocking" interchangeably, as they share the same goal but are complementary. 
In this paper, we adhere to the definition provided by~\cite{DeepBlocking} and focus  on the blocking phrase. 

Thus, our focus is on tackling the problem of efficiently blocking tables $\E$ in order to generate 
$\kw{Ca}(\E)$.
Our objective is to develop effective solutions that expedite the blocking procedure.

\eat{
Much of the literatures uses two concepts, data partitioning and blocking,  interchangeably, due to these two 
these two phrase share the same goal but are complementary.
In this paper, we follow the definition of literature~\cite{DeepBlocking} and focus on the blocking phrase.
That is, we consider the problem of blocking  tables $\E$ to produce set of all
pairs of matching entity profiles $\D(\E) = \{ (t, s):t\in \E , s\in \E, t = s\}$..
We seek to develop solutions that speed up the blocking procedure. 
}
}

\stitle{Rule-based blocking.}
We study rule-based blocking in this paper,
due to its efficiency and explainability remarked earlier.
We review a class of matching dependencies (\MDs), originally proposed in~\cite{fan2011dynamic}. \looseness=-1

\etitle{Predicates.}
Predicates over schema $R$ are defined as follows:
\vspace{-0.7ex}
\[ 
    p::= t.A=c\ |\ t.A=s.B\ |\  t.A \approx s.B
\vspace{-0.7ex}
\]
where  $t$ and $s$ are tuple variables denoting tuples of $R$, $A$ and $B$ are attributes
of $R$ and $c$ is a constant; 
$t.A = s.B$ and $t.A=c$ compare the equality on \emph{compatible} values,
\eg $t.\eid = s.\eid$ says that $(t,s)$
\revise{is a potential match};
$t.A \approx s.B$ 
compares the \emph{similarity} of $t.A$ and $s.B$.
\reviseX{Here any similarity measure, 
 symmetric or asymmetric, can be used as $\approx$, 
\eg 
edit distance or KL divergence,} such that $t.A \approx s.B$  is true
if $t.A$ and $s.B$ are “similar” enough w.r.t. a threshold.
\reviseS{Sophisticated similarity measures like ML models can also be used as in~\cite{REE-discovery,rock}.}
\looseness=-1

\etitle{Rules.}
A (bi-variable) {\em matching dependency} (\MD) over $R$ is:
\vspace{-0.7ex}
\[\varphi = X \ra l,
\vspace{-0.7ex}
\]
where  $X$ is a conjunction of predicates over $R$
with two tuple variables $t$ and $s$, and $l$ is
$t.\eid = s.\eid$.
We refer to $X$ as the {\em precondition}
of $\varphi$, and $l$ as the {\em consequence} of $\varphi$, respectively.

\begin{example}\label{example:2}
Consider a (simplified)  e-commence database with self-explained schema
\kw{Products}(\kw{eid}, \kw{pno}, \kw{pname}, \kw{price}, \kw{sname} (store name), \kw{description}, \kw{color}, \kw{saddress} (store address)).
Below are some examples \MDs, where the rule in Example~\ref{example:1} is written as $\varphi_1$.

\sstab
(1) $\varphi_1:  t.\kw{color} = s.\kw{color}  \land t.\kw{price} = s.\kw{price} \land t.\kw{sname} = s.\kw{sname} \allowbreak \land t.\kw{pname} \approx_\kw{ED} s.\kw{pname} \to t.\kw{eid} = s.\kw{eid}$,
where $\approx_\kw{ED}$ measures the 
 edit distance.
\revise{As stated before, $\varphi_1$ identifies two products, by their
colors, prices, product names and the stores sold. }\looseness=-1

\sstab
(2) $\varphi_2:  t.\kw{sname} = s.\kw{sname} \land t.\kw{description} \approx_\kw{JD} s.\kw{description} \to t.\kw{eid} \allowbreak = s.\kw{eid}$,
where $\approx_\kw{JD}$ measures the 
 Jaccard distance.
 The \MD says that if two products are sold in the store and have a similar description, then they are identified as a potential match.

\sstab 
(3) $\varphi_3: t.\kw{saddress} \approx_\kw{ED} s.\kw{saddress} \land t.\kw{description} \approx_\kw{JD} s.\kw{description} \allowbreak  \to t.\kw{eid} = s.\kw{eid}$.
It gives another condition for identifying two products,
\ie the two products with similar descriptions sold from stores with similar addresses 
are potentially matched.
\end{example}
\looseness=-1

\vspace{-0.6ex}
\etitle{Semantics}. 
A {\em valuation} of tuple variables of an \MD
$\varphi$ in $D$, or simply
{\em a valuation of $\varphi$},
is a mapping $h$ that instantiates the two variables $t$ and $s$ with
tuples in $D$.
A valuation $h$ {\em satisfies} a predicate $p$ over $R$,
written as $h \models p$, if the following is
satisfied: (1) if $p$ is $t.A = c$ or  $t.A = s.B$, then it is interpreted
as in tuple relational calculus following the standard
semantics of first-order logic \cite{AbHuVi1995};
and (2) if $p$ is $t.A \approx s.B$, then $h(t).A \approx h(s).B$
returns true.
Given a conjunction $X$ of predicates, we say $h \models X$ if for
{\em all} predicates $p$ in $X$, $h \models p$.

\etitle{Blocking.}
\revise{Rule-based blocking 
employs a set  $\Delta$ of \MDs.}
Given a partition $P$,
a pair $(t_1, t_2) \in P \times P$ is in $\kw{Ca}(P)$ \emph{iff}
there exists an \MD $\varphi$ in $\Delta$ 
such that the valuation $h(t_1, t_2)$ of $\varphi$ that instantiates variables $t$ and $s$ with tuples $t_1$ and $t_2$ satisfies the precondition of $\varphi$;
we call such  $\varphi$ as a \emph{witness} at $(t_1, t_2)$,
since it indicates that $(t_1, t_2)$ is a potential match.
Otherwise,
$(t_1, t_2)$ will be filtered.  
\reviseS{Since a precondition is a conjunction of predicates,
rule-based blocking is in \emph{Disjunctive normal form} (DNF), \ie 
it is to evaluate a disjunction of conjunctions.}
\looseness=-1

\begin{example}\label{example:3}
Continuing with Example~\ref{example:2},
consider $D$ in Table~\ref{tab:products}
and $h(t_1, t_4)$ that instantiates variables $t$ and $s$ with tuples $t_1$ and $t_4$ in $D$.
Since $h(t_1, t_4)$ satisfies the precondition of $\varphi_1$, $\varphi_1$ is a witness at $(t_1, t_4)$.
Similarly, one can verify that $\varphi_1$ is not a witness at $(t_2, t_3)$.
\eat{As an example, consider \MDs: $\varphi_1$ and $\varphi_2$.
Our goal is to identify potential matches between pairs of math in the dataset based on a self-explained schema (\kw{Products(pname, price, sname, desc, color)} as shown in Table~\ref{tab:products}).
We have three pairs of valuations: ($t_1$.\kw{eid} =$t_4$.\kw{eid}), ($t_4$.\kw{eid} =$t_5$.\kw{eid}), and ($t_2$.\kw{eid}= $t_3$.\kw{eid}). 
The first two satisfy $\varphi_1$, while the last one satisfies $\varphi_2$. }
\end{example}
\looseness=-1

\vspace{-0.6ex}
\etitle{Discovery of \MDs.}
\reviseY{\MDs can be considered as a special
case of entity enhancing rules (\REEs)~\cite{REE-discovery, topk}.
We can readily \revise{apply} the discovery algorithms for \REEs, \eg~\cite{REE-discovery, topk}, to discover \MDs (details omitted). }
\looseness=-1

\vspace{-0.5ex}
\stitle{GPU hardware.}
\reviseY{As general processors for high-performance computation,
GPUs offer the following benefits compared with CPUs.} 

\reviseX{First, GPUs provide massive  parallelism
by programming with CUDA (Compute Unified Device Architecture)~\cite{CUDA}. 
A GPU has multiple SMs (Streaming Multiprocessors),
where each SM accommodates multiple processing units.
\eg V100 has 80 SMs, each with 64 CUDA cores. 
SMs handle the parallel execution of CUDA cores.
In CUDA programming, CUDA cores are conceptually organized into TBs~(Thread Blocks) and physically grouped into thread warps, each comprising subgroups of 32 threads.
This hierarchical organization allows 
thousands of threads running simultaneously on GPUs. 
} \looseness=-1

\eat{
subgroups of 32 threads, called warps, are executed in parallel.
TBs~(Thread Blocks) used to organize and execute parallel tasks on GPUs. 
This organization leads to a hierarchical parallel model: within each TB, subgroups of 32 threads, called warps, are executed in parallel.
A thread block consists of a group of threads that collaborate and execute concurrently within a Streaming Multiprocessor (SM). 
This organizational structure allows for efficient data sharing through shared memory and synchronization via barrier instructions like __syncthreads(). Thread blocks play a crucial role in optimizing parallel computations by leveraging the massive parallelism offered by GPUs.

This organization leads to a hierarchical parallel model: within each TB, subgroups of 32 threads, called warps, are executed in parallel.}

\reviseX{
Second, GPUs utilize the DMA (Direct Memory Access) technology, which enables direct data transfer between GPU memory and system memory. 
This not only reduces CPU overhead but also allows the GPU to handle multiple data streams simultaneously.
However, the number of PCIe lanes determines the maximum number of streams that can transfer data simultaneously (\eg 16 PCIe lanes for V100). 
When multiple partitions perform data transfers over a PCIe lane, only one can utilize the lane at a time. 
}

\eat{Secondly, the communication interface between the GPU and the rest of the system is provided by PCIe lanes.
The number of PCIe lanes determines the maximum number of streams that can transfer data simultaneously (\eg Gen3$\times$16 PCIe lanes for V100). 
When multiple partitions perform data transfer over a PCIe lane, only one can utilize the lane, while the others must wait.
} 
\looseness=-1

\eat{ 
Secondly,
GPUs are equipped with wider/faster memory interfaces, enabling quick access to large data. Using tightly integrated high-bandwidth memory technologies, GPUs provide a device memory with bandwidth approaching 1 TB/sec. 
This is particularly useful for analytic tasks with data-intensive operations.}

\eat{
GPUs offer a variety
of features, like programmable on-chip shared memory, similar to a CPU cache, with a peak bandwidth of
several TB/sec, thread and warp synchronization primitives, and designated techniques to overlap computation with I/O. }

\reviseX{
Third, GPUs adopt \emph{SIMT (Single Instruction, Multiple Threads)} execution, where 
each SIMT lane is an individual unit
that is responsible for executing a thread under a single instruction.
\emph{Thread divergence} can adversely affect the performance
and it typically occurs in conditional statements (\eg \emph{if-else}),
where some lanes take one execution path while the others take a different path. 
However, 
GPUs must execute different execution paths sequentially, rather than in parallel, 
resulting in 
underutilization of GPU resources.
}
\looseness=-1

\eat{
One key difference between a GPU and a CPU is that the GPU adopts SIMT (Single Instruction, Multiple Threads) execution.  
Each SIMT lane, individual processing units within a SIMT, processes one element of a vector of data under the control of a single instruction.
Analyzing the utilization of SIMT lanes is crucial for determining whether a program fully exploits the massive parallelism of the GPU.
\emph{SIMT divergence} can adversely affect performance and typically occurs in \emph{if-else} blocks of GPU kernels. 
Some lanes may need to execute the \emph{if} block while others may need to execute the \emph{else} block. However, because the same instruction must be executed in all lanes, the runtime will process each \emph{if} block and each \emph{else} block sequentially, resulting in sub-optimal utilization of processing units.
Therefore, tuning blocking to be GPU-conscious is a non-trivial task that requires exploring such unique characters of GPU.
}

\section{{HyperBlocker}: System Overview}
\label{sec-overview}

In this section, we present the overview of \kw{HyperBlocker},
\revise{a  GPU-accelerated system for rule-based blocking} that 
optimizes the efficiency by considering  rules, 
underlying data, and hardware simultaneously. 
In the
literature, GPUs and CPUs are usually referred to as
devices and hosts, respectively. We also follow this
terminology.  


\vspace{-0.7ex}
\stitle{Challenges.}
Existing parallel blocking methods typically rely on multiple CPU-powered machines under the shared nothing architecture, to achieve data partition-based parallelism.
They reduce the runtime by using more machines,
which, however, is not always feasible due to, \eg the increasing communication cost \reviseC{(see Section~\ref{sec-intro})}. 
\eat{Moreover,
\revise{it is
hard 
to strike for} a good balance between parallelism and \revise{recall}.
Although
a more fine-grained data partitioning strategy can improve parallelism,
it inevitably affects the \revise{recall} since 
a match $(t_1, t_2)$ may be missed if $t_1$ and $t_2$ are distributed to different machines.} 
\looseness=-1

\reviseC{In light of these, \Hyper 
focuses on parallel blocking under a shared memory architecture;
this
introduces new challenges.} 

\eat{
While existing parallel blocking has been commonly conducted on the CPUs architecture,
rule-based blocking  under a hybrid CPUs/GPUs architecture  
presents new challenges.
}

\vspace{-0.3ex}
\etitle{(1) \revise{Execution plan for efficient blocking}.}
The efficiency of blocking depends heavily on how
much/fast we can filter mismatches.
Therefore, a good execution plan that specifies how rules are evaluated is crucial.
\reviseS{However, most existing blocking optimizers 
fail to consider the properties of rules/data  for blocking and even the optimizers of popular DBMS (\eg PostgreSQL~\cite{postgresql}) 
may not work well when handling similarity comparisons and evaluating queries in DNF (see Section~\ref{sec-plan}).}
\eat{For example, PostgreSQL's optimizer,  
popular advanced open source relational database, 
uses a naive strategy to optimize execution plan for \MDs, 
contributing to inefficient execution for blocking~(see Section~\ref{sec-plan}).}
This motivates us to design a different 
\reviseC{plan generator}.

\vspace{-0.4ex}
\etitle{(2) Hardware-aware parallelism.}
When GPUs are involved, \reviseC{those CPU-based techniques adopted in existing solvers} no longer suffice,
\eat{
For instance, a GPU consists of multiple streaming multiprocessors (SM). 
Each SM accommodates multiple vector units.
This hardware organization results in a hierarchical parallel
model: each CUDA kernel includes groups of threads called
thread blocks~(TBs). Within each TB, subgroups of threads called warps are executed
simultaneously. }
since  GPUs have radically different characteristics~(Section~\ref{sec-background}). 
\reviseX{Novel GPU-based parallelism for blocking is required to improve GPU utilization, \eg by reducing thread wait stalls and thread divergence.
} \looseness=-1


\vspace{-0.3ex}
\etitle{(3) Multi-GPUs collaboration.}
Existing parallel \reviseX{blocking} solvers  focus on minimizing the communication cost across all workers~\cite{dis-dup,deng2022deep}.
\reviseC{
However, this objective no longer applies in multi-GPUs scenarios, where unique challenges such as task decomposition,
(inter-GPU) resource management and task scheduling arise.}

\vspace{-0.7ex}
\stitle{Novelty.}
The ultimate goal of \Hyper is to generate the set $\kw{Ca}(P)$ of potential matches on each data partition $P$.
To achieve this, we implement three novel components as follows:

\vspace{-0.4ex}
\etitle{(1) Execution plan generator (EPG) (Section~\ref{sec-plan}).}
We develop a generator
to generate data-aware and rule-aware execution plans,
\reviseS{which support arbitrary comparisons and work well with DNF evaluation.}
\revise{Here
we say an execution plan is data-aware,
since it considers the distribution of data to decide which predicates are evaluated first; 
similarly, it is rule-aware,
since it is optimized for underlying rules. }
\looseness=-1


\vspace{-0.3ex}
\etitle{(2) Parallelism optimizer (Section~\ref{sec-exec-model}).}
We implement a specialized optimizer that exploits the hierarchical structure of GPUs,
to optimize the power of GPUs by utilizing thread blocks~(TBs) and warps. 
With this optimizer, \reviseX{
blocking can be effectively parallelized on GPUs}. 


\vspace{-0.3ex}
\etitle{(3) Resource scheduler (Section~\ref{sec-exec-model}).} 
To achieve optimal performance \reviseY{over multiple GPUs,}
\reviseY{a partitioning strategy} and 
a resource scheduler are developed to manage the resources, 
 and balance the workload
across multiple GPUs,
minimizing idle time and resource waste.

\eat{ 
Efficient resource scheduling is vital to ensure optimal utilization of the GPU's capabilities. 
\Hyper incorporates heuristic resource scheduling techniques that 
enable workload balancing across multiple GPUs 
and efficient management of GPU resources. 
This ensures that GPU resources are utilized effectively, 
minimizing idle time and maximizing overall system performance.
}

\begin{figure}[t]
	\centering
	\includegraphics[width=0.76\linewidth]{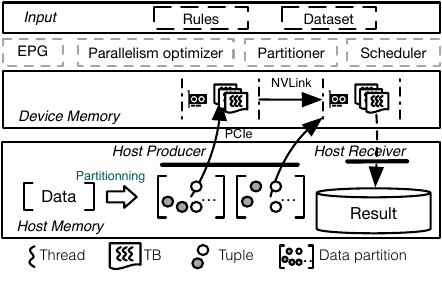}
        \vspace{-5ex}
	\caption{\revise{The pipelined architecture of \kw{HyperBlocker}}}
	\label{fig:pipeline}
\vspace{-3ex}
\end{figure} 

\vspace{-0.8ex}
\stitle{Architecture.}
The architecture of \Hyper is shown in Figure~\ref{fig:pipeline}.
Taken a relation $D$ of tuples and 
a set $\Delta$ of \MDs discovered offline as input,
\Hyper divides the tuples in $D$ into $m$ disjoint partitions
and asynchronously processes partitions in a pipelined manner,
so that the execution at devices and the data transfer can be overlapped,
\CR{mitigating the excessive data transfer costs.}
\looseness=-1

\vspace{-0.3ex}
\etitle{Workflow.}
More specifically, 
\Hyper works in five steps:

\vspace{-0.5ex}
\revise{ 
\sstab
(1) 
\textit{Data partitioning.}
\Hyper divides the tuples in $D$ into $m$ partitions,
to allow parallel processing asynchronously.
}

\vspace{-0.3ex}
\sstab
(2) \textit{Execution plan generation.}
Given the set $\Delta$ of \MDs discovered offline,
an execution plan that specifies in what order the rules (and the predicates in rules) should be evaluated is generated at the host.  
\looseness=-1

\vspace{-0.3ex}
\sstab
(3) 
\textit{Host scheduling.}
The blocking on each partition forms a computational task
\revise{and
the host dynamically assigns} tasks to the queue(s) of available devices without interrupting their ongoing execution, \reviseY{minimizing the idle time of devices and improving resource utilization.} \looseness=-1

\vspace{-0.3ex}
\sstab
(4) 
\textit{\reviseX{Device} execution.}
When a device receives the task assigned,
it conducts the rule-based blocking on the corresponding data partition, following the execution plan generated in Step (2).

\vspace{-0.3ex}
\sstab
(5) 
\textit{Result retrieval.}
Once a task is completed on a device, 
the host will pull/collect the result (i.e., $\kw{Ca}(P)$) from the device. 

\vspace{0.6ex}
To facilitate processing, \Hyper has two additional components: \kw{HostProducer} and \revise{\kw{HostReceiver}},
where the former manages Steps (1), (2), and (3) and the latter handles Step (5). 
Steps (3) (4) (5) in \Hyper work asynchronously in a pipeline manner.
\section{EPG: Execution Plan Generator}
\label{sec-plan}

Given the set $\Delta$ of \MDs and a partition $P$ of $D$,
\revise{a naive approach to compute $\kw{Ca}(P)$ is to evaluate each \MD in $\Delta$ for all pairs in $P$.}
That is,
to decide whether $(t_1, t_2)$ is in $\kw{Ca}(P)$,
we perform $O(\sum_{\varphi\in\Delta}|\varphi|)$ predicate evaluation,
where $|\varphi|$ is the number of predicates in $\varphi$.  
Worse still,
there are $O(|\Delta|!|\varphi|!)$ possible ways to evaluate all \MDs in $\Delta$,
since both \MDs in $\Delta$ and predicates in each $\varphi$ can be evaluated in arbitrary orders.
However, not all orders are equally efficient.

\looseness=-1

\vspace{-0.3ex}
\begin{example}
	In Example~\ref{example:3}, 
	$\varphi_1$ is a witness at $(t_1, t_4)$
	while $\varphi_3$ is not.
	If we first evaluate $\varphi_1$ for $(t_1, t_4)$, 
	$(t_1, t_4)$ is identified as a potential match and there is no need to evaluate $\varphi_3$. 
	Moreover, 
	when evaluating $\varphi_1$ for another pair $(t_2, t_3)$,
	we can conclude that 
	$\varphi_1$ is not a witness at $(t_2, t_3)$,
	as soon as \revise{we find $h(t_2, t_3) \not \models t.\kw{price} = s.\kw{price}$.} 
	\eat{let's explore the evaluation of $\varphi_1$ further. If we first evaluate the predicate $t$.\kw{color} = $s$.\kw{color}, we might find that all tuples satisfy this condition. In this case, we would need to evaluate the subsequent predicate for the entire set of tuple pairs.
		However, if we evaluate $t$.\kw{owner} = $s$.\kw{owner} first, we can quickly discard tuples, \eg ($t_1$, $t_2$), that are guaranteed to not match, executing comparisons only between the rest.}
	\label{exa:order}
\end{example}
\looseness=-1

\vspace{-1.5ex}
\stitle{Challenges.}
Given the huge number of \revise{possible evaluation} orders,
it is non-trivial to define \revise{a good one}, for three reasons:

\reviseX{\etitle{(1) Rule priority.}
Recall that \reviseS{rule-based blocking is in DNF, \ie}
as long as there exists a witness at $(t_1, t_2)$,
$(t_1, t_2)$ will be considered as a potential match.
This motivates us to prioritize the rules in $\Delta$ 
so that promising ones can be evaluated early;
once a witness is found, the evaluation of the remaining rules can be skipped. \looseness=-1

\vspace{-0.1ex}
\etitle{(2) Reusing computation.} 
\MDs may have common predicates.
To avoid evaluating a predicate repeatedly,
we 
reuse previous results whenever possible,
\eg
given $(t_1, t_2)$, $\varphi_1: p \land X_1 \ra l$ and $\varphi_2: p \land   X_2 \allowbreak \ra  l$,
if $\varphi_1$ is not a witness at $(t_1, t_2)$ since $h(t_1, t_2) \not \models p$,
neither is $\varphi_2$. 
\looseness=-1

\vspace{-0.1ex}
\etitle{(3) Predicate ordering.}
Given $(t_1, t_2)$ and $\varphi: X \ra l$, $\varphi$ is not a witness at $(t_1, t_2)$ 
if we find the first $p$ in $X$ such that  
$h(t_1, t_2) \not \models p$.
\eat{In other words, if $(t_1, t_2)$ fails to satisfy prior predicates in $\varphi$, the remaining evaluation of $\varphi$ can be bypassed directly.
While this is undeniably obvious, ``many approaches have not leveraged it effectively''~\cite{li2023selection}.}
However,
to decide which predicate is evaluated first,
we have to consider both its evaluation cost and its effectiveness/selectivity.
}
\looseness=-1

\reviseS{
\vspace{0.36ex}
As remarked in~\cite{preedet,rock},
blocking with a set of \MDs can be implemented in a single DNF SQL query (\ie an {\tt OR} of 
{\tt ANDs}),
where similarity predicates in \MDs are re-written as user-defined functions (UFDs).
In light of this,
one may want to adopt the optimizers of existing DBMS to tackle rule-based blocking,
which, however, may not work well, for several reasons.
\eat{(a) The compilation/translation of UDFs  may introduce significant overheads.
As evidenced in~\cite{rock},
even modern engines like \kw{SparkSQL}~\cite{sparksql}and \kw{Presto}~\cite{prestosql} cannot finish the execution in
one day when SQLs are embedded with complicated UDFs.}
(a) The mixture of relational operators and UDFs poses serious challenges to an optimizer~\cite{rheinlander2017optimization}.
It may lack ``the information needed to decide whether they can be reordered with
relational operators and other UDFs''~\cite{hueske2013peeking}
and worse still,
it is hard to accurately estimate the runtime performance of UDFs~\cite{rheinlander2017optimization}.
(b) Using {\tt OR} operators in {\tt WHERE} clauses can be inefficient, since it can force the database to perform a full table scan to find matching tuples~\cite{badOR}. 
(c) Similar to~\cite{li2023selection}, 
if a tuple pair fails to satisfy prior predicates in a blocking rule, the remaining evaluation of this rule can be bypassed directly.
While this is undeniably obvious, ``many approaches have not leveraged it effectively''~\cite{li2023selection}.

\eat{ 
To justify, 
we tested \kw{PostgreSQL}~\cite{postgresql}, a popular DBMS, on different queries,
by (a) specifying similarity predicates in different orders 
and (b) evaluating SQLs with or without {\tt OR} operators (see details in~\cite{full}).
An in-depth analysis using \kw{PostgreSQL}'s {\tt EXPLAIN} feature reveals the following: 
(a) Although \kw{PostgreSQL} may prioritize equality, 
it does not optimize the order of similarity predicates well,
\eg it evaluates similarity predicates simply as specified in the SQLs, leading to 20\% slowdown on average.
(b) When handling {\tt OR} operators, \kw{PostgreSQL} may perform sequential scans with nested loops, without utilizing the hash index.
These findings highlight the need of designated optimization for rule-based blocking,
where we have to handle similarity comparisons and DNF evaluation efficiently. \looseness=-1
}

}

\vspace{-0.5ex}
\stitle{Novelty.}
In light of these, 
\revise{EPG in \Hyper gives a lightweight solution, by generating an execution plan} 
to make the overall evaluation cost of $\Delta$ as small as possible.
Its novelty includes
(a) a new notion of execution tree \reviseS{that works no matter what types of comparisons are used},
(b) a rule-aware scoring strategy, to decide which \MDs in $\Delta$ are evaluated first, and
(c) a data-aware predicate ordering scheme, to strike for a balance between cost and effectiveness.

\vspace{0.3ex}
Below we first give the formal definition of execution plans
and then show how EPG generates a good execution plan.

\vspace{-1ex}
\subsection{Execution plan}
An execution plan specifies how rules and predicates in $\Delta$ are evaluated.
Although an
	execution plan can be represented in different ways, we represent it as an \emph{execution tree}, denoted by $\T$ in this paper for its conciseness and simplicity 
\reviseX{(an example is given in Figure~\ref{fig:tree}, to be explained in more detail later)}.
(1) A node in $\T$ is denoted by $N$,
where the root is denoted by $N_0$.
(2) A \emph{path $\rho$ from the root}
is a list $\rho = (N_0, N_1, \ldots, N_L)$
such that $(N_{i-1}, N_i)$ is an edge of $\T$ for $i\in[1, L]$;
the length of $\rho$ is $L$, \ie the number of edges on $\rho$.
(3) We refer to $N_2$ as a \emph{child node} of $N_1$ if $(N_1, N_2)$ is an edge in $\T$, and as a \emph{descendant} of $N_1$ if there exists a path from $N_1$ to $N_2$;
conversely,
we refer to $N_1$ as a \emph{parent node} (resp. predecessor) of $N_2$.
(4) Each edge $e$ represents a predicate $p$
and is associated with a score, denoted by $\kw{score}(e)$,
indicating the priority of $e$.
(5) A node is called a \emph{leaf} if it has no children and $\T$ has $|\Delta|$ leaves,
where each leaf is associated with a rule  $\varphi: X \ra l$ in $\Delta$;
the length of the path from the root to the leaf is $|X|$ (\ie the number of predicates in $X$)
and 
for each predicate in $X$, it appears exactly once in an edge on the path.
(6) The leaves of two \MDs may have common predecessors, in addition to the root;
intuitively,
this means that the \MDs have common predicates.
With a slight abuse of notation, we also denote an execution plan by $\T$.
\looseness=-1

\vspace{-0.5ex}
\stitle{Evaluating an execution plan.}
For each pair $(t_1, t_2)$,
it is evaluated by exploring  $\T$ via depth-first search (DFS), starting at the root.
At each internal node $N$ of $\T$,
we pick a child $N_c$
such that the edge $(N, N_c)$,
\reviseY{whose associated predicate is $p$,}
has the highest score among all children of $N$.
Then we check whether  $h(t_1, t_2) \models p$.
If it is the case,
we move to $N_c$ and process $N_c$ similarly.
Otherwise,
we check whether $N$ still has other unexplored children and we process them similarly,
according to the decreasing order of scores.
If all children of $N$ are explored,
we return to the parent $N_p$ of $N$
and repeat the process.
The evaluation completes if we reach \reviseS{the first} leaf of $\T$.
Suppose the rule associated with this leaf is $\varphi: X \ra l$.
This means  $h(t_1, t_2)$ satisfies all predicates in $X$, along the path from the root to that leaf,
and thus $h(t_1, t_2) \models X$,
\ie\ 
\reviseS{we find a witness at $(t_1, t_2)$,
and the remaining tree traversal can be skipped.}
\looseness=-1

\begin{example}
\vspace{-0.5ex}
	\label{exa:DFS}
	\eat{Give an execution tree with previous \MDs}
	Consider an execution tree $\T$ in Figure~\ref{fig:tree}(b),
	which depicts \MDs in Example~\ref{example:2}.
	For simplicity, we denote a predicate $t.A = s.A$ (resp. $t.A \approx s.A$) by $p_A^=$ (resp. $p_A^\approx$) and the score associated with each edge is labeled.
	DFS starts at the root, which has two children.
	It first explores the edge labeled $p_\kw{sname}^=$
	since its  score is higher.
	When DFS completes, \MDs $\varphi_2$, $\varphi_1$ and $\varphi_3$ are checked in order.
\vspace{-1.5ex}
\end{example}
\looseness=-1

\vspace{-1ex}
\subsection{Execution plan generation}
\label{subsec:generation}
Taking
the set $\Delta$ of \MDs as input,
EPG in \Hyper returns an execution plan $\T$ in the following two major steps:

\sstab
(1) We order all predicates appeared in $\Delta$,
by estimating their evaluation costs via a shallow model and quantifying their probabilities of being satisfied,
by investigating the underlying data distribution.
\looseness=-1

\sstab
(2) Based on the predicate ordering,
we build an execution tree $\T$ by iterating \MDs in $\Delta$.
Moreover,
we compute a score for each edge in $\T$,
by considering 
the probability of finding a witness, \ie reaching a leaf, \revise{if we explore $\T$ following this edge.}

\vspace{0.6ex}
Note that 
plan generation in EPG can be regarded as a \reviseC{(quick)} pre-processing step for blocking,
\ie
once an execution plan is generated,
it is applied in \emph{all} partitions of $D$.
Below we present these two steps.
For simplicity,
we assume \kwlog that 
$D$ is itself a partition. 
\looseness=-1

\vspace{-0.5ex}
\stitle{Predicate ordering.}
Denote by $\P$ the set of all predicates appeared in $\Delta$.
Intuitively, 
not all predicates in $\P$ are equally potent for evaluation,
\eg although texts (\eg\ \kw{description}) are
often more informative than categorical attributes (\eg\ \kw{color}), 
the former comparison is more expensive.
\reviseX{A simple idea is to order predicates by only considering attribute types and operators 
\reviseS{(\eg prioritize equality like traditional optimizers)}.
However, the time/effect of evaluating a predicate for distinct tuples can be different.
Without taking the underlying data into account,
it can lead to poor ordering.
Motivated by this, we order the predicates in  $\P$ by their ``cost-effectiveness'' on~$D$.
\eat{Similar trade-off also appears in other scenarios, 
\eg query optimization in RDBMS~\cite{moerkotte2006building}.}}
\looseness=-1

For simplicity, below we consider a predicate $p$ that compares $A$-values of two tuples,
\ie $t.A = s.A$ or $t.A \approx s.A$ (simply $p_A^=$ or $p_A^\approx$).
\revise{All discussion extends to other predicate types,
	\reviseY{\eg $t.A = s.B$.}
 } \looseness=-1

\etitle{Evaluation cost.}
We measure the evaluation cost of a predicate $p$ by the time for evaluating $p$;
a predicate that can be evaluated quickly should be checked first.  
Given a predicate $p$ in $\P$ and a relation $D$,
the evaluation cost of $p$ on $D$,
denoted by $\kw{cost}(p, D)$, is:
\vspace{-0.7ex}
\[
\kw{cost}(p, D) = \sum_{(t_1, t_2) \in D \times D} T_p(t_1, t_2).
\vspace{-0.7ex}
\]
where $T_p(t_1, t_2)$
denotes the \reviseY{actual time} for
checking $h(t_1, t_2) \models~p$. 

Note that 
it can be costly to iterate all tuple pairs in $D$ \CR{to compute the exact evaluation cost of $p$ on $D$.
\eat{ 
\eg on average,
it takes more than 100s to compute $\kw{cost}(p, D)$ on a dataset with 2M tuples (see more in~\cite{full}).}
Thus,} below we train a shallow NNs, denoted by $\N$, (\ie a small feed-forward neural network~\cite{kraska2018case}) to estimate the exact $T_p(t_1, t_2)$, since it has been proven effective in approximating a continuous function on a closed interval \cite{Stone_Weierstrass_theorem}.
\looseness=-1

\eetitle{Shallow NNs.}
The inputs of $\N$ are two tuples $t_1$ and $t_2$, and a predicate $p$, that compares the $A$-values of $t_1$ and $t_2$.
It first encodes the attribute type and the $A$-value of $t_1$ into an embedding $\vec{t_1}$; similarly for $\vec{t_2}$.
The embeddings are then fed to a feed-forward neural network,
which outputs the estimated time for evaluating $p$ on $(t_1, t_2)$. \looseness=-1
\reviseX{We train $\N$ offline with training data sampled from historical logs,
so that the training data follows the same distribution as $D$\eat{  (see~\cite{full})}.}

\eat{
To train $\N$,
we collect training data from historical logs,
where
each training instance is in form $(x, y)$;
here $x$ is triplet  $(p, t_1, t_2)$ and $y$ is its label, \ie the measured time of evaluating $p$ at $(t_1, t_2)$.
We train $\N$ offline, using stochastic gradient descent with the mean square error loss.
While the training time is not our focus,
	$\N$ converges quickly in a few pass~\cite{kraska2018case},
	since it has few parameters.}

\eetitle{Estimated cost.}
\revise{Based on $\N$,
	the estimated cost of $p$ on $D$
	is}
\vspace{-0.7ex}
\[
\hat{\kw{cost}}(p, D) = \kw{norm}\big(\sum_{(t_1, t_2) \in D \times D} \N(p, t_1, t_2)\big),
\vspace{-0.7ex}
\]
\revise{where $\kw{norm}(\cdot)$ normalizes the estimated cost in the range (0,1].}
\CR{In practice, we can also use a  sampled set from $D$ for estimating costs.}
\looseness=-1

\eat{
\eetitle{Remark.}
As will be verified in Section~\ref{sec-expt}, although $\hat{\kw{cost}}(p,D)$ is computed by iterating tuple pairs in $D$ (\ie quadratic cost),
it is much faster than computing $\kw{cost}(p, D)$.
Better still, 
the predicate orderings derived from $\hat{\kw{cost}}(p, D)$ is close to that derived from $\kw{cost}(p, D)$. \looseness=-1

}

\eat{
\eetitle{Remark.}
Although $\hat{\kw{cost}}(p,D)$ seems to require quadratic evaluation, the prediction costs are very low, as \kw{EPG} evaluate the pruning cost of predicates using a few sampled tuple pairs from the test data, rather than all pairs. 
}

\etitle{Effectiveness.}
We measure the \reviseC{effectiveness} of predicate $p$
by \reviseC{its selectivity,} \ie the probability of being satisfied.
Given the attribute $A$ compared in $p$, 
we quantify how likely $t_1$ and $t_2$ have distinct/dissimilar values on $A$.
If $t_1$ and $t_2$ \reviseY{do so with a high probability,}
$p$ is less likely to be satisfied;
such predicate should be evaluated first since it concludes that an \MD involving $p$ is not a witness early. \looseness=-1

To achieve this,
we investigate the data distribution in $D$.
Specifically,
we use LSH~\cite{LSH} to hash the $A$-values of all tuples into $k$ buckets,
so that similar/same values are hashed into the same bucket with a high probability,
where $k$ is a predefined parameter. 

Denote the number of tuples hashed to the $i$-th bucket by $b_i$.
Intuitively,
the evenness of hashing results reflects the probability of $p$ being satisfied.
If all tuples are hashed into the same bucket,
it means that the $A$-values of all tuples are similar
and thus $p$ (which compares the $A$-values) is likely to be satisfied
by many pairs $(t_1, t_2)$;
such predicates should be evaluated with low-priority.
%
Motivated by this,
the probability of $p$ being satisfied on $D$, denoted by $\kw{sp}(p, D)$,
is estimated by measuring the evenness of hashing,
\ie  \looseness=-1
\vspace{-0.7ex}
\[
\kw{sp}(p, D) = \kw{norm}\big(\sqrt{ \frac{1}{k} \sum_{i=1}^k (b_i - \frac{|D|}{k})^2}\big)
\vspace{-0.7ex}
\]

\begin{figure}[t]
	\centering	\includegraphics[width=0.95\linewidth]{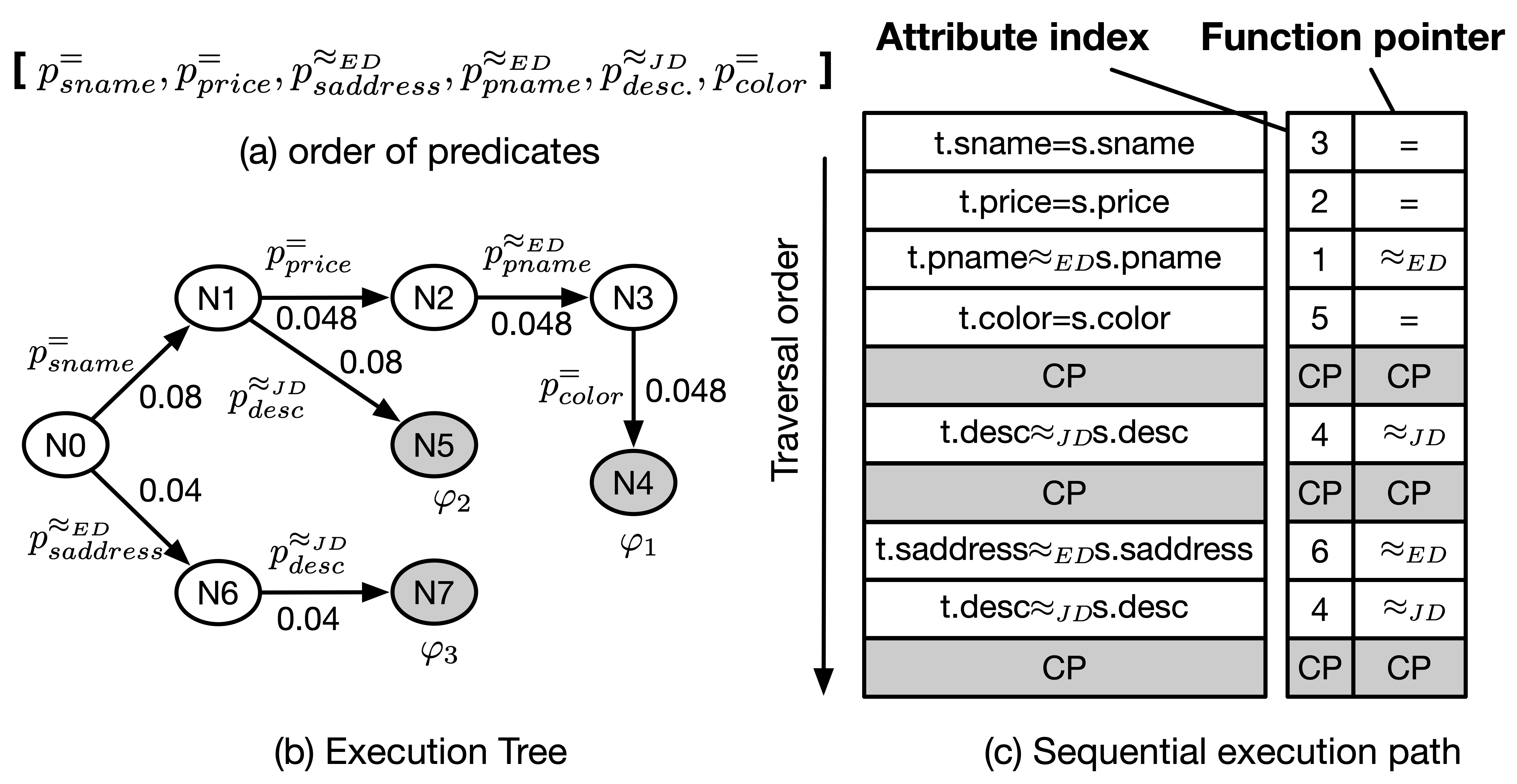}
	\vspace{-3.5ex}
	\caption{Execution tree}
	\label{fig:tree}
	\vspace{-3ex}
\end{figure} 

\etitle{Ordering scheme.}
Putting these together,
we can order all the predicates $p$ in $\P$ by the cost-effectiveness, defined to be $\frac{1- \kw{sp}(p, D)}{\hat{\kw{cost}}(p, D)}$.
Intuitively,
hard-to-satisfied predicates will be evaluated first,
since they are more likely to fail a rule,
while costly predicates will be penalized,
to strike a balance between the cost and the effectiveness. \looseness=-1

\begin{example}
\vspace{-0.5ex}
\label{exa:equality}
	\eat{Give examples of cost, sp, and orders}
	Consider two predicates $p_\kw{color}^=$ and $p_\kw{pname}^{\approx_\kw{ED}}$ in $\varphi_1$.
	On the one hand, since $p_\kw{color}^=$ is an equality comparison while $p_\kw{pname}^{\approx_\kw{ED}}$ computes the edit distance,
    $p_\kw{pname}^{\approx_\kw{ED}}$  is more costly to evaluate, \eg $\hat{\kw{cost}}(p_\kw{color}^=, \allowbreak D) = 0.1 < \hat{\kw{cost}}(p_\kw{pname}^{\approx_\kw{ED}}, D) = 0.6$.
	On the other hand,
	since all tuples in $D$ have the same color
	(and satisfy $p_\kw{color}^=$),
	we have $\kw{sp}(p_\kw{color}^=, D)=1$;
	similarly, let $\kw{sp}(p_\kw{pname}^{\approx_\kw{ED}}, D)=$ 0.4.
	Then
	the cost-effectiveness of $p_\kw{color}^=$  and $p_\kw{pname}^{\approx_\kw{ED}}$
	are $\frac{1-1}{0.1} = 0$ and $\frac{1-0.2}{1} = 0.8$, respectively,
	and $p_\kw{pname}^{\approx_\kw{ED}}$ is ordered before  $p_\kw{color}^=$ (see Figure~\ref{fig:tree}(a)).
\end{example}
\vspace{-1.5ex}
\looseness=-1

\stitle{Constructing an execution tree.}
\revise{We initialize the execution tree $\T$ with a single root node $N_0$.
	Then based on the predicate ordering,
	we progressively construct $\T$} by processing the \MDs in $\Delta$ one by one. 
For each \MD $\varphi: X \ra l \in \Delta$,
we assume the predicates in $X$ are sorted in the descending order of their cost-effectiveness,
\ie if $X $ is $ p_1 \land p_2 \land \ldots \land p_{|X|}$,
then $\frac{1- \kw{sp}(p_i, D)}{\hat{\kw{cost}}(p_i, D)} > \frac{1- \kw{sp}(p_j, D)}{\hat{\kw{cost}}(p_j, D)}$ for $1 \leq i < j \leq |X|$.
We traverse $\T$, starting from the root, and process the predicates in $X$, starting from $p_1$.
Suppose that the traversal is at a node $N$ 
and the predicate we are processing is $p_i$.
We check the children of $N$. 
If there exists a child node $N_c$ of $N$ such that the edge $(N, N_c)$ represents $p_i$, we move to this child and process the next predicate $p_{i+1}$ in $X$.
Otherwise, we create a new child node $N_c$  for $N$ such that the edge $(N, N_c)$ represents $p_i$, move to this new child and process the next predicate $p_{i+1}$ in $X$.
The traversal process continues until all predicates in $X$ are processed
and we set the current node we reach as a leaf node,
whose associated rule is $\varphi$.
\looseness=-1

\begin{example}
	\eat{An example of tree construction}
	The predicate ordering is shown in Figure~\ref{fig:tree}(a).
	Assume that we have processed $\varphi_1$ and created path $(N_0, N_1, \allowbreak N_2, N_3, N_4)$ in $\T$ in Figure~\ref{fig:tree}(b).
	\revise{Then we show how $\varphi_2: p_\kw{sname}^{=} \land \allowbreak p_\kw{description}^{\approx_\kw{JD}} \ra l$ is processed.}
	We start from the root and process $p_\kw{sname}^{=}$.
	Since there is a child $N_1$ of root labeled $p_\kw{sname}^{=}$,
	we move to $N_1$ and process $p_\kw{description}^{\approx_\kw{JD}}$.
	Since there is no child of $N_1$ labeled $p_\kw{description}^{\approx_\kw{JD}}$,
	we create a new $N_5$ and label $(N_1, N_5)$ as $p_\kw{description}^{\approx_\kw{JD}}$.
	Since all predicates in $\varphi_2$ are processed,
	$N_5$ is a leaf node, whose associated rule is $\varphi_2$.
\end{example}
\looseness=-1

Intuitively, given $(t_1, t_2)$ and $\varphi \in \Delta$,
if $\varphi$ is more likely to be a witness at $(t_1, t_2)$,
it should be evaluated earlier.
Motivated by this,
we compute the probability for $\varphi: X \ra l$ to be a witness on $D$ as: \looseness=-1

\vspace{-1.5ex}
\[
\kw{wp}(\varphi, D) = \prod_{p \in X} \kw{sp}(p, D)
\vspace{-0.7ex}
\]
if we assume the satisfaction of predicates as independent events;
intuitively,
if all predicates in $X$ are satisfied,
$\varphi$ is a witness.
If this does not hold,
\reviseY{we can reuse historical logs and estimate $\kw{wp}(\varphi, D)$,
to be the proportion of historical pairs such that $\varphi$ is a witness.}
Since the evaluation of \MDs in $\Delta$
is guided by edge scores during DFS on $\T$,
below we define the score of a given edge $e$ based on $\kw{wp}(\varphi, D)$. \looseness=-1

\etitle{Edge score.}
For each \MD $\varphi$,
we denote  by  $\rho_\varphi$ the path of $\T$ from root to the leaf whose associated \MD is $\varphi$.
We compute the set of \MDs $\varphi$ 
in $\Delta$ such that the given edge $e$ is part of $\rho_\varphi$
and denote it by
$\Psi_e$,
\ie $\Psi_e = \{\varphi \in \Delta \mid e $ is part of $ \rho_\varphi\}$.
The score of edge $e$ is 
$\kw{score}(e) = \max_{\rho_\varphi \in \Psi_e} \kw{wp}(\varphi, D)$.
This said,
edges leading to promising \MDs will have high scores and thus, will be explored early via DFS on $\T$. \looseness=-1

\begin{example}
	\eat{An example of edge scores}
	Let  $\kw{sp}(p_\kw{sname}^=, D) = 0.4$ and 
	$\kw{sp}(p_\kw{description}^{\approx_\kw{JD}},  D) = 0.2$.
	Then $\kw{wp}(\varphi_2, D) = 0.4 \times 0.2 = 0.08$.
	Assume that
	we also compute  $\kw{wp}(\varphi_1, D) = 0.048$.
	Then the score of edge $e = (N_0, N_1)$ is 
	$\max\{ \allowbreak \kw{wp}(\varphi_1, D), \allowbreak \kw{wp}(\varphi_2, D)\}$ = 0.08,
	since $e$ is part of both $\rho_{\varphi_1}$ and $\rho_{\varphi_2}$.
\end{example}
\looseness=-1

\revise{ 
	\vspace{-0.8ex}
	\stitle{Complexity.}
	It takes EPG $O(c_\kw{unit}|\P| + |\P|\log(|\P|)+|\varphi||\Delta|)$ time to generate the execution plan,
	where $c_\kw{unit}$ is the unit time for computing the cost-effectiveness of a predicate.
	This is because the predicate ordering can be obtained in $O(c_\kw{unit}|\P| + |\P|\log(|\P|))$ time
	and the tree can be constructed in $(|\varphi||\Delta|)$ time,
	by scanning  $\Delta$ once.
	\looseness=-1
	
}

\vspace{-0.4ex}
\stitle{Remark.}
 \reviseY{As a by-product of ensuring the predicate ordering and  DFS tree traversal, 
we can reuse the evaluation results of common ``prefix'' predicates
(\ie common predecessors in $\T$).}
\reviseS{Moreover,
if a tuple pair fails to satisfy the predicate associated with edge $(N, N_c)$  in $\T$,
the evaluation of all descendants of $N_c$ is bypassed directly.}
\looseness=-1

\begin{example}
	We evaluate $\T$ in Figure~\ref{fig:tree}(b) for $(t_1, t_5)$ in $D$.
	After evaluating $p_\kw{sname}^=$,
	we find that $h(t_1, t_5) \allowbreak \not \models p_\kw{description}^{\approx_\kw{JD}}$ and thus we cannot move to $N_5$.
	Then DFS will return back to $N_1$
	and continue to check unexplored children of $N_1$ (\ie $N_2$).
	In this way, the common ``prefix'' predicate $p_\kw{sname}^=$ of $\varphi_1$ and $\varphi_2$ is only evaluated once.
\vspace{-1ex}
\end{example}
\looseness=-1
	\vspace{-0.5ex}
\section{Optimizations and Scheduling}
\label{sec-exec-model}
\reviseX{As remarked earlier,
GPUs adopt SIMT execution,
where a thread is idle if other threads take longer
(\ie thread divergence). 
Below are sources of divergence (some are specific to rule-based blocking). \looseness=-1

\mbi 
\item \textit{Conditional statements.}
GPUs may execute different paths in conditional statements (Section~\ref{sec-background}),
\eg one pair may be quickly identified as a potential match if the first \MD checked is its witness,
while another is found as a mismatch until all \MDs are iterated. \looseness=-1
\item \textit{Data-dependent execution.}
The execution depends on the data being processed,
\eg
 even for the same predicate, the evaluation time on different tuples is different (\eg long vs. short text). 
\item \textit{Imbalanced workloads.}
If the workload assigned to each thread is not evenly distributed, some may complete faster than others.
\mei

While thread divergence is a general issue in GPU-programming, rule-based blocking offers some unique opportunities to mitigate it,
\eg\ \reviseS{the evaluations of distinct pairs are often \emph{independent} tasks,} making it possible to (a) assign approximately equal tasks to threads, to enable \emph{workload balancing},
and 
(b)
``steal'' tasks from other threads,
to cope with different \emph{execution paths}
and \emph{data-dependent execution}.} 
Below
we present the hardware-aware optimization and scheduling techniques that exploit GPU characteristics for massive parallelism,
\reviseC{including:} (a) efficient \reviseX{device} execution of an execution plan (Section~\ref{subsec:plan}),
\reviseY{(b) strategies to mitigate divergence}
(Section~\ref{subsec:balance}),
\eat{(c) a strategy to avoid parallel write conflicts (Section~\ref{subsec:conflict}),}
(c) collaboration of multiple GPUs (Section~\ref{subsec:GPUs}).
\eat{and more in~\cite{full}.}
\eat{More optimization strategies are presented in \cite{full}.}
\looseness=-1

\vspace{-1ex}
\subsection{Execution plan on GPUs}
\label{subsec:plan}
The execution plan $\T$, initially generated on CPUs, will undergo the evaluation on GPUs in a DFS manner. 
\reviseX{
However, DFS tree traversal is typically recursively implemented, which is not efficient on GPUs.
It may exacerbate divergence since 
each call adds a recursive function to the stack and incurs message payloads (see Section~\ref{sec-expt}). 
\eat{Code transformation may help alleviate such issues.}}
Moreover,
although we can reuse ``prefix'' predicates via DFS,
some predicates may still be evaluated repeatedly,
\eg $p_\kw{description}^{\approx_\kw{JD}}$ in $\varphi_2$ and $\varphi_3$.
Optimized structures are required 
to harness the power of GPUs. 

\eat{
However, tree traversal on GPUs is not efficient due to 
the thread divergence (resulting from the branch divergence).
Moreover,
although we can reuse the results of ``prefix'' predicates via DFS on $\T$,
some predicates may still be evaluated repeatedly,
\eg $p_\kw{description}^{\approx_\kw{JD}}$ in $\varphi_2$ and $\varphi_3$.
Optimized data structures are required to harness the processing power of GPUs. \looseness=-1
}

\vspace{-0.6ex}
\stitle{Tree traversal on GPUs.}
Note that upon completion of the tree construction, 
the evaluation order is fixed.
Thus the DFS traversal of the tree on CPUs
can be translated to a \emph{sequential execution path}, which is an ordered list of predicates, on GPUs
(see Figure~\ref{fig:tree}(c) for the sequential execution path of the tree in Example~\ref{exa:DFS}).

We maintain two structures for each predicate $p$ in the execution path: an index buffer and a function pointer buffer,
which 
store the indices of attributes compared in $p$, and the function pointer of the comparison operator in $p$, respectively,
\eg
for predicate $p_\kw{sname}^=: t.\kw{sname} = s.\kw{sname}$,
its comparison operator is ``='' and its attribute index is 3 since \kw{sname} is the 3rd attribute in schema \kw{Products}.
In addition, at the end of each rule, we set a checkpoint (\kw{CP}).
When a GPU thread encounters a \kw{CP}, 
it knows that the undergoing tuple pair satisfies a rule and it can
skip the subsequent computation. \looseness=-1

\vspace{-0.6ex}
\stitle{Reusing computation.}
To avoid repeated evaluation,
we additionally maintain a bitmap for all predicates on GPUs.
The bit of a predicate $p$ is set to true if $p$ has been evaluated.
If this is the case,
we can directly reuse previous results. 
\reviseX{This bitmap can also be used for symmetric predicates (\ie $h(t_1, t_2) \models p$ iff $h(t_2, t_1) \models p$).}

\reviseX{Note that to be general,
we do not make the assumption that a witness $\varphi$ at $(t_1, t_2)$
is also a witness at  $(t_2, t_1)$,
due to, \eg asymmetric similarity comparison (Section~\ref{sec-background}).
However, we can extend \Hyper if such assumption holds,
by maintaining a bitmap to avoid  repeated evaluation for $(t_2, t_1)$
if $(t_1, t_2)$ is already evaluated.
}

\vspace{-1ex}
\subsection{\reviseY{Divergence mitigation strategies}}
\label{subsec:balance}

\reviseY{To further mitigate divergence,} we propose two 
GPU-oriented strategies, namely \emph{parallel sliding windows (PSW)} and \emph{task-stealing}.

\vspace{-0.6ex}
\stitle{Parallel sliding windows (PSW).}
Given a partition $P$, 
PSW processes it with only a few index jumps;
it also helps GPUs evenly distribute workloads across SMs.
\reviseY{Specifically, 
PSW works in 3 steps:}

\vspace{-0.2ex}
\sstab
(1) We divide $P$ into $\frac{|P|}{n_t}$ intervals,
where each interval consists of $n_t$ tuples.
These intervals are processed with a fixed-size window, which slides the intervals from left to right. 
\reviseY{Within each window,
we assign an interval to a Thread Block (TB) with 
warps of 32 threads; 
each thread in the TB 
is responsible for a tuple $t_i$ in the interval.}
\looseness=-1

\vspace{-0.2ex}
\sstab
(2) Assume that a thread is responsible for tuple $t_i$. 
Then this thread compares $t_i$ \reviseY{with all the other tuples, say $t_j$,} in $P$ according to the execution plan $\T$ and decides whether  $(t_i, t_j)$ is a potential match.  

\vspace{-0.2ex}
\sstab
(3) When all threads of a TB finish,
this TB writes the results back to the host memory
and 
it will move on to process the next interval in the next sliding window \reviseY{until the window reaches the end.}

\eat{ 
In the PSW method, the entity set, 
denoted as $\E$, is divided into $\frac{|\E|}{32}$ disjoint intervals. 
Then, $P$ TBs are allocated to the first $P$ intervals, 
where $P$ is the number of SMs. 
Each TB comprises 32 threads, 
and each thread is assigned an entity of the given interval. 
Next, for each TB, it executes plan $\P$ on the corresponding execution interval and $\E$. 
After going through all the pairs of execution intervals and $\E$, the TB writes the resulting data back to the host memory and moves on to the next interval, continuing this process until all intervals have been covered. In total, for each TB, PSW requires only $\frac{|\E|}{32P}$ sequential index jumps to process each interval.
}

\reviseY{Note that in total,}
it requires $\frac{|P|}{n_tn_w}$ sequential index jumps for each TB,
where $n_w$ is the size of the sliding window.

\begin{figure}[t]
	\centering
	\includegraphics[width=0.95\linewidth]{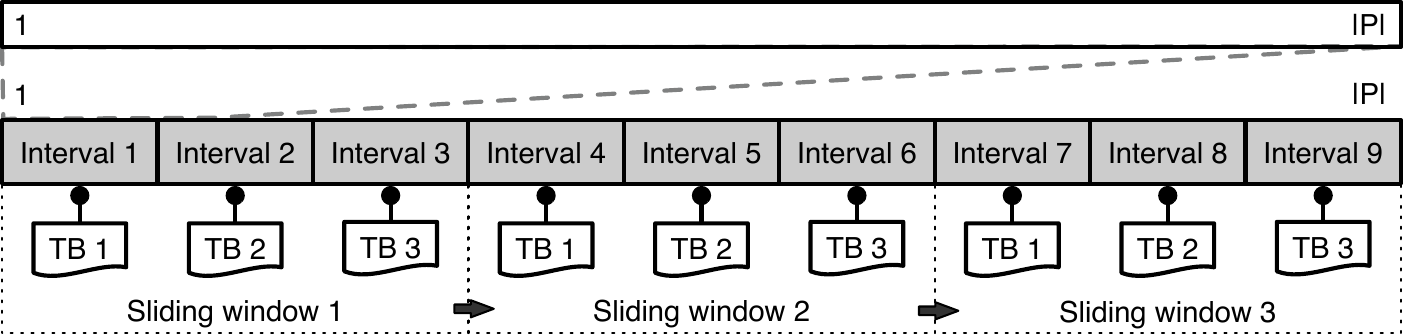}
 \vspace{-2.5ex} 
	\caption{Parallel sliding windows\eat{The tuples of $\E$ are divided into intervals. 
	Each interval is assigned a TB,
	and Cartesian product with $\E$.}}
	\label{fig:sliding_window}
\vspace{-4ex} 
\end{figure}

\begin{example}
\label{exa:PSW}
As shown in Figure~\ref{fig:sliding_window}, a data partition $P$ is divided 9 intervals and the size of the sliding window is 3 (\ie $n_w = 3$).
Interval 1 is assigned to TB1,
where each thread in TB1 will compare a tuple in Interval 1
with all other tuples in $P$.
When all threads of TB1 finish evaluation,
TB1 moves on to process Interval 4.
\end{example}

\vspace{-1ex}
\stitle{Task-stealing.}
\reviseX{Although
each TB will process roughly equal intervals,
the execution time of different intervals is not the same, due to conditional statements and data-dependent execution remarked earlier.}
This said, the workloads of all TBs can still be imbalanced. 

\begin{figure}[t]
	\centering
	\includegraphics[width=0.95\linewidth]{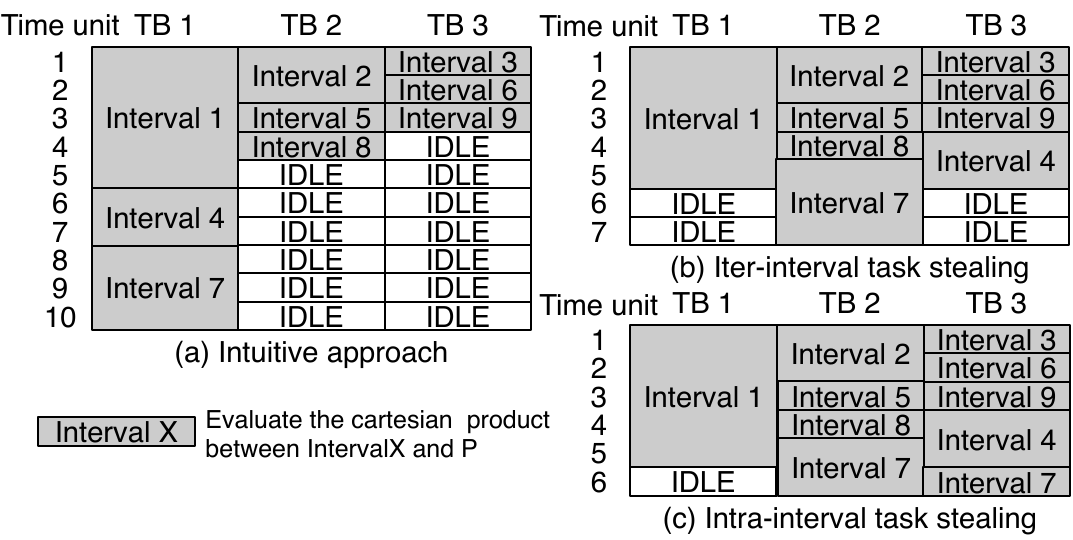}
\vspace{-3.5ex}
	\caption{Task-stealing\eat{Work flow of three approaches based on an example \warn{(Connect previous Fig, Step to Time unit, part (b) adds a bitmap, part (c) TB3 steals Interval 7)}}}
	\label{fig:task_stealing}
\vspace{-3ex}
\end{figure}

\begin{example}
\label{exa:steal}
Continuing Example~\ref{exa:PSW},
\reviseY{three TBs process 9 intervals in Figure~\ref{fig:task_stealing}(a).
Even though each TB is assigned 3 intervals,}
the execution times can still skew,
\eg 
the total time units required by TB1 and TB3 
are 10 and 3, respectively,
\reviseY{\ie TB3 is idle for 7 time units.}
\end{example}
\looseness=-1

Below we introduce both the inter-interval and intra-interval task-stealing strategies to further balance the workloads.

\etitle{Inter-interval task-stealing}.
It is commonly observed that
the execution times of some TBs are longer than the others.
In this case,
a large number of TBs are idle, waiting for the slowest TB. 

In light of this,
we employ an inter-interval task-stealing strategy.
\reviseY{Specifically,
we maintain a bitmap in global memory, where each bit indicates the status of an interval,}  
so that TBs can steal not-yet-processed intervals from each other.
Each TB processes intervals in two stages:
(a)
It first processes its assigned intervals one by one.
\reviseY{Whenever a TB starts to process an interval, 
the bitmap is checked.
If the bit of the interval is false (\ie not yet processed),}
\revise{it processes this interval and sets the bit true.}
(b) If this TB is idle after finishing all assigned intervals,
it traverses the bitmap to steal a not-yet-processed interval,
by setting the corresponding bit true and processing that interval.
\revise{Other TBs will skip an interval if it has been stolen.}
\looseness=-1

\eat{ 
This is actually a common occurrence, when the distribution 
\wrt the length of the entities of an interval held by each TBs is skewed, 
a large number of threads are idle, waiting for the large interval to finish. 
We thus devise the Task-Stealing strategy to make the idle threads work on the unprocessed Invervals. 
To do this we fist introduce a bitmap datastructure, 
that is a sequence of bits (length=$\frac{|\E|}{32}$), each of which indicates the
status of a Interval. Set operations on bitmap are
very simple and efficient.
It is store at the shared memory,
so that other TBs can steal the workloads 
With bitmap, we schedule the processing of residual intervals in two stages. 
In the first stage, all TBs process their own interval until one finishes.
Whenever an Interval starts to be processed, 
the corresponding TB set the value in to be true.
In the second stage, idle TBs traverse bitmap to get an unprocessed Interval.
The remaining TBs will skip to processed invervals that have been stolen.}

\begin{example}
\label{exa:steal-inter}
In Example~\ref{exa:steal}, 
TB3 finishes its assigned intervals after 3 time units.
Then it checks the bitmap and steals \reviseY{Interval 4};
similarly for TB2.
Compared with the time in Figure~\ref{fig:task_stealing}(a), 
the total time units are reduced from 10 to 7 after stealing in Figure~\ref{fig:task_stealing}(b).
\end{example}

\vspace{-1ex}
\etitle{Intra-interval task-stealing.}
\reviseY{Recall that a thread for $t_i$ will compare $t_i$ with other tuples in $P$.}
Since the evaluation of distinct pairs
is independent,
\reviseY{we can even} steal tasks
from executing intervals. 
To facilitate this,
we maintain two integers \kw{start} and \kw{end}, initialized to 1 and $|P|$, respectively,
\reviseX{indicating the remaining range of tuples to be compared with $t_i$.}
Then this thread \revise{starts to evaluate $(t_i, t_{\kw{start}})$.
Upon completion,}
it sets $\kw{start} =  \kw{start} +1 $
and moves on to the next pair $(t_i, \allowbreak t_{\kw{start}})$.
When $\kw{start} = \kw{end}$,
this thread finishes all evaluation for $t_i$. \looseness=-1

Based on this, the intra-interval task-stealing works as follows.
If TB$_a$ finishes all assigned intervals and 
there are no not-yet-processed intervals,
it finds an executing TB$_b$ and iterates all threads in TB$_b$,
so that the $i$-th thread in TB$_a$ steals half workload (\ie half pairs to be compared)
from the $i$-th thread in TB$_b$.
Assume the integers maintained for the $i$-th thread in 
TB$_b$ (resp. TB$_a$) are $\kw{start}_b$ and $\kw{end}_b$ (resp. $\kw{start}_a$ and $\kw{end}_a$).
We set 
$\kw{start}_a = \kw{start}_b + \frac{\kw{start}_b + \kw{end}_b}{2}$,
$\kw{end}_a = \kw{end}_b$,
and 
$\kw{end}_b = \allowbreak \kw{start}_b + \frac{\kw{start}_b + \kw{end}_b}{2} - 1$,
\reviseX{\ie the latter half of tuples remained to be compared is stolen from each thread in 
TB$_b$.} \looseness=-1

\begin{example}
Continuing Example~\ref{exa:steal-inter}, when TB3 finishes Interval 4 stolen from TB1 
in Figure~\ref{fig:task_stealing}(b),
it finds no not-yet-processed intervals.
However,
since TB2 is still evaluating Interval 7,
TB3 steals half remaining workload from it, saving 1 more time unit  (Figure~\ref{fig:task_stealing}(c)).
\end{example}
\looseness=-1

\vspace{-1.8ex}
\subsection{GPU collaboration}
\label{subsec:GPUs}
A GPU server nowadays usually has multiple GPUs connected via NVLink~\cite{nvlink} or PCIe.
Scaling blocking to multiple GPUs is beneficial for jointly utilizing the computation and storage powers of GPUs.

In pursuit of this, one can split data evenly
so that each GPU handles exactly one~\cite{even-spliting},
or assign multiple partitions to each GPU in a round-robin manner~\cite{round-robin}.
These, however, do not work well
since (a) workload can be imbalanced due to skewed execution times of partitions\eat{ (see statistic in~\cite{full})},
\eat{(b)  \reviseY{pending partitions may} wait when multiple partitions compete for the same \reviseX{PCIe lane} (\eg \reviseX{16 lanes} for V100 GPU)}
\reviseX{(b) pending partitions may wait when multiple partitions compete for limited PCIe bandwidth~or CUDA cores 
(see Section~\ref{sec-background})}
and (c) they independently conduct blocking on partitions and do not effectively handle scenarios where $t_i$ and $t_j$ reside on different partitions, resulting in elevated false-negative rates. 
\reviseY{To address these, 
one can duplicate tuples in multiple partitions~\cite{deng2022deep, dis-dup},
but it incurs both memory and data transfer costs.}


In light of these, we present a collaborative approach integrating partitioning and scheduling strategies,
where the former aims at minimizing data redundancy while reducing false negatives
and the latter prioritizes load balancing and minimizes resource contention.

\vspace{-0.8ex}
\stitle{Data partitioning.}
\reviseY{A typical method for data partitioning
computes a hash key} for each tuple based on some attributes and tuples with same hash key are grouped together.
Instead of sacrificing the accuracy (\eg using only one hash function) or 
unnecessarily duplicating tuples, 
\Hyper applies $s$ hash functions to obtain
$s$ partition-keys, where $s$ is the number of children $N_c$ of the root node $N_0$ in the execution tree $\T$;
each hash function is constructed from the predicate $p$ associated with an edge $(N_0, N_c)$.
\reviseS{In this way, the predicates that we adopt for  data partitioning are those prioritized by $\T$,
\eg 
given $p_\kw{sname}^=$ associated with ($N_0, N_1$) in Figure~\ref{fig:tree}(b),
we hash tuples in $D$ based on their values in \kw{sname}.}
The benefits are two-fold: 
(1) According to the construction of $\T$,
these hash functions are \reviseC{selective} and might be shared by rules,
\ie we can achieve good hashing with a few hashing functions.
(2)
We can assign 
each tuple a branch ID, indicating the hash function used. 
Only tuples that share the same hash function
are compared, thereby reducing redundant computations incurred by multiple hash functions.
\looseness=-1

\vspace{-0.3ex}
\stitle{Scheduling.}
\Hyper adopts a two-step scheduling strategy. 
Initially, data partitions and GPUs are hashed to random locations on a unit circle~\cite{CHBL}. If a partition $P_i$ is assigned to an \emph{ineligible} GPU \reviseX{(where there is no idle core or available PCIe bandwidth),} it is rerouted to the nearest available GPU in a clockwise direction.

\etitle{Remark.}
If data partitioning is done by a hashing function from a similarity predicate $p$,
it is possible that $h(t_1, t_2) \models p$ but $t_1$ and $t_2$ reside on different partitions, leading to potential false negatives in blocking.
In this case,
\reviseY{a CUDA kernel~\cite{CUDA_guide}}
with local data $P_i$  can optionally ``pull'' partition $P_j$ from another kernel and evaluate $\T$ across $P_i$ and $P_j$. 
The pull operation retrieves data from locations outside $P_i$, depending on whether $P_i$ and $P_j$ reside on the same GPU. If  $P_i$ and $P_j$ reside on the same GPU, the pull operation is executed directly without any data transfer. Otherwise, the pull operation for $P_j$ can be carried out using {\tt cudaMemcpyPeer()} to take the advantages of high bandwidth and low latency provided by NVLink.
\looseness=-1

\section{Experimental Study}
\label{sec-expt}
We evaluated \Hyper for its accuracy-efficiency and
scalability. 
\reviseC{We also conducted sensitivity tests and 
ablation studies.}

\vspace{-0.7ex}
\stitle{Experimental setup.} 
We start with the experimental setting.

\begin{table}[t]
        \setlength{\tabcolsep}{2.6pt}
	\caption{\reviseX{Datasets}}\label{tab:dataset}
        \vspace{-2ex}
	\centering
        \fontsize{5.8pt}{6.5pt}\selectfont 
	\begin{tabular}{ccccccccc}
		\toprule  
		\textbf{Dataset}      & \textbf{Domain}  & \textbf{\#Tuples } & \textbf{Max \#Pairs} & \textbf{\#GT Pairs} & \textbf{\#Attrs} & \textbf{\#Rules} & \textbf{\#Partitions} \\ 
		\midrule
        \revise{\kw{Fodors}-\kw{Zagat}} &restaurant     &     866   & $1.8\times 10^4$    &     112       &    6 & 1 & 1 \\
		\kw{DBLP}-\kw{ACM} &citation     &     4591   & $6.0\times 10^6$    &     2294       &    4 & 10 & \revise{8} \\ 
        \kw{DBLP}-\kw{Scholar} &citation     &     66881   & $1.7\times 10^8$   &     5348       &    4  &10 & \reviseY{8} \\
		\kw{IMDB} &movie     &    1.5M   & $8.1\times 10^{10}$    &      0.2M+        &    6 &10 &128\\ 
		\kw{Songs}		 &music    &     0.5M     &     $2.7\times 10^{11}$  &     1.2M   & 8 &10 & \reviseY{128}\\ 	
		\kw{NCV} &vote     &     2M   & $1.0\times 10^{12}$    &       0.5M+      &   5   &10 & 512\\ 
		\kw{TFACC} &traffic     &  10M   & \reviseY{$1.0\times 10^{14}$}    &       \#       &     16 &50 & 1024\\
        \revise{\kw{TFACC_{large}} }&    traffic     &  36M   & $1.3\times 10^{15}$   &       \# &  16 &50 & 1024\\
		\bottomrule
	\end{tabular}	
\vspace{-3ex}
\end{table}

\etitle{Datasets.}
We used \revise{eight} real-world public datasets in \reftab{tab:dataset},
which are widely adopted ER benchmarks and real-life datasets~\cite{magellandata,MOT,Dedoop_dataset}. 
\reviseY{\reviseC{Most} datasets (except \kw{TFACC} and $\kw{TFACC}_\kw{large}$) have
labeled matches or mismatches as the ground truths (GT).
For datasets without ground truths, we assume the original datasets were correct, 
and randomly duplicated tuples as noises~\cite{preedet}. }
The training data 
consists of 50\% of ground truths and  50\% of randomly selected noise.

\etitle{Baselines.}
\revise{As remarked in Section~\ref{sec-background},  
although \Hyper is designed as a blocker,
it can be used with or without a matcher.
Thus,
below we not only compared \Hyper against widely used blockers
but also integrated ER solutions (\ie blocker + matcher).}
\looseness=-1

We compared three distributed ER systems: (1) \kw{Dedoop}\cite{dedoop,code_dedoop}, (2) \kw{SparkER}\cite{sparkER,code_sparker}, (3) \kw{DisDedup}~\cite{dis-dup,code_disdedup}, where
\kw{DisDedup} is  the SOTA CPU-based parallel ER system, designed to minimize communication and computation costs; \kw{Dedoop} focuses on optimizing computation cost;
\kw{SparkER} integrates Blast blocking \cite{BLAST} on Spark~\cite{spark}. \looseness=-1

We also compared
four GPU-based baselines:
\revise{
(4) \kw{DeepBlocker} \cite{DeepBlocking},
}
(5) \kw{GPUDet}~\cite{GPU-ER}, (6) \kw{Ditto}~\cite{ditto,code_ditto}, \revise{(7) $\kw{DeepBlocker}_\kw{Ditto}$, 
where \kw{DeepBlocker} is the SOTA DL-based blocker,} 
\kw{GPUDet} implements well-known similarity algorithms for tuple pair comparison, 
\reviseY{\kw{Ditto} is the SOTA matcher,} 
\revise{and 
$\kw{DeepBlocker}_\kw{Ditto}$  uses \kw{DeepBlocker} as the blocker and \kw{Ditto} as the matcher, respectively.}
Note that \kw{Ditto} takes tuple pairs as input, instead of \reviseY{relations/partitions} as other methods.
Due to the high cost of \kw{Ditto},
it is infeasible to feed the Cartesian product of data to \kw{Ditto}.
Thus, for each tuple in \CR{GT},
we adopted a similarity-join method \eat{\kw{Faiss}}~\cite{Faiss} to get the top-2 nearest neighbors, as \CR{its preprocessing step.}
\revise{Denote the resulting baseline by $\kw{Ditto}_\kw{top2}$}.
\eat{Since \kw{Faiss} often serves as a key component of many ER solutions~\cite{DeepER,DeepBlocking},
\reviseX{we also compared \kw{Faiss} in~\cite{full}.}}
\looseness=-1

\eat{
Among the above baselines,
\kw{Ditto} is typically used as the accurate matcher in the final matching phase of ER,
while the others can be used in ER, with or without a matcher.
}

\eat{ 
\etitle{Data partitioning}\label{subsec:blocking}.
We used suffix array data partitioning~\cite{blocking-filtering-survey} as the default partitioning method. 
It converts each key of a tuple into a list of its suffixes that are longer than a predetermined minimum length.  
Then, it defines a partition key for every suffix by a random hash function. 
Tuples sharing the same partition key are grouped together.
}

Besides, \reviseY{we also implemented several variants:} 
(1) \Hyper, the basic blocker with all optimizations. 
(2) $\Hyper_\kw{Ditto}$, an improved version  that uses \Hyper as the blocker and \kw{Ditto} as the matcher, respectively. 
\revise{Note that $\Hyper_\kw{Ditto}$ is particularly compared against $\kw{Ditto}_\kw{top2}$ to show
how we speed up the overall ER.}
(3) $\Hyper_\kw{noEPG}$, a variant 
without EPG (Section~\ref{sec-plan}).
(4) $\Hyper_\kw{noHO}$ that disables all hardware optimizations (Section~\ref{sec-exec-model}).
\reviseX{We also compared more designated variants in Exp 3-5.}
\looseness=-1

\begin{table}[t]
\centering
\setlength{\tabcolsep}{3.1pt}
\caption{\reviseX{Comparison with the SOTA DL-based blocker}}
		\label{table:dl_vs_rule}
\vspace{-2ex}
\fontsize{5.8pt}{6.5pt}\selectfont 
\begin{tabular}{ccccc}
\toprule
\multirow{2}{*}{\textbf{Method}} & \multirow{2}{*}{\textbf{Metric}} & \multicolumn{3}{c}{\textbf{Dataset}}                                            \\ \cline{3-5} 
                                 &                                  & \textbf{Fodors-Zagat} \hspace{2ex} & \textbf{DBLP-Scholar} \hspace{2ex} & \textbf{DBLP-ACM} \\ \hline

\multirow{4}{*}{\kw{DeepBlocker}}& \kw{Rec}~($\%$) &  100~(+0) & 98~(+5) & 98~(+4) \\

                                 & \kw{CSSR}~(\textpertenthousand) &  15.1~(+14.5) & 2.3~(+1.1) & 2.2~(+1.8) \\
                                 
                                 & Time~(s) & 6.1~(122$\times$) & 72.8~(11.0$\times$) & 8.0~(10.0$\times$) \\ 
                                 & \tabincell{c}{Host Mem. cost~(GB)} & 9.9~(49.5$\times$) & 14.0~(23.3$\times$) & 10.3~(34.3$\times$) \\ 
                                 & \tabincell{c}{\reviseX{Device} Mem. cost~(GB)} & 0.9~(1.8$\times$) & 1.1~(1.6$\times$) & 0.9~(1.5$\times$) \\
                                 \hline
                                 
\multirow{4}{*}{\Hyper}          & \kw{Rec}~($\%$) & 100 & 93 & 94 \\
                                 & \kw{CSSR}~(\textpertenthousand) & 0.6 & 1.2 & 0.4\\
                                 
                                 & Time~(s) & 0.05 & 6.6 & 0.8 \\ 
                                 & \tabincell{c}{Host Mem. cost~(GB)} & 0.2 & 0.6 & 0.3 \\ 
                                 & \tabincell{c}{\reviseX{Device} Mem. cost~(GB)} & 0.5 & 0.7 & 0.6\\  
\bottomrule
\end{tabular}
\vspace{-3ex}
\end{table}

\vspace{-0.4ex}
\etitle{Rules.}\label{subsec:rule}
We mined  \MDs\ 
\reviseY{using~\cite{rule_mining} and the number of \MDs is shown in Table~\ref{tab:dataset}.}
We checked 
the \MDs manually to ensure correctness.
\eat{The number of \MDs adopted for each dataset is shown in Table~\ref{tab:dataset}.}
\looseness=-1

\begin{table*}[]
        \caption{Accuracy \& runtime on benchmarks  where ``*''  denotes that integrating  \Hyper with \kw{Ditto} does not improve the F1-score and thus we report the result of \Hyper, and ``/'' denotes that the F1-score cannot be computed 
        within 3 hours.
        } \label{tab:overview}
        \vspace{-2ex}
	\begin{minipage}[t]{1\linewidth}
		\centering
        \setlength{\tabcolsep}{5pt}
		{\fontsize{6.7pt}{1em} 
                \selectfont 
			\begin{tabular}{c cc cc cc cc cc}				
                \toprule				
                \multirow{2}{*}{\textbf{Method}} & \multirow{2}{*}{\textbf{Backend}} & \multirow{2}{*}{\textbf{Category}}      & \multicolumn{2}{c}{\textbf{DBLP-ACM}}       & \multicolumn{2}{c}{\textbf{IMDB}}           & \multicolumn{2}{c}{\textbf{Songs}}          & \multicolumn{2}{c}{\textbf{NCV}}            \\ 
                
                \cmidrule(lr){4-5} \cmidrule(lr){6-7} \cmidrule(lr){8-9} \cmidrule(lr){10-11}
				   &    &    & \multicolumn{1}{c}{\textbf{F1-score}} & \textbf{Time~(s)} & \multicolumn{1}{c}{\textbf{F1-score}} & \textbf{Time~(s)} & \multicolumn{1}{c}{\textbf{F1-score}} & \textbf{Time~(s)} & \multicolumn{1}{c}{\textbf{F1-score}} & \textbf{Time~(s)} \\ 
       
        		\midrule
                
                \kw{SparkER} & CPU & Blocker & 0.77~(-0.17) & 11.0~(13.8$\times$) & 0.31~(-0.65)  & 242.9~(6.8$\times$) & 0.08~(-0.72) & 203.4~(15.2$\times$) & 0.26~(-0.66) & 229.3~(49.8$\times$) \\
                \kw{GPUDet} & GPU & Blocker  & 0.92~(-0.02) & 20.1~(25.1$\times$) & 0.94~(-0.02) & 323.8~(9.1$\times$) & 0.80~(+0) & 404.8~(30.2$\times$) & 0.90~(-0.02) & 1252.6~(272.3$\times$) \\        
                \kw{DeepBlocker} & GPU & Blocker  & 0.98~(+0.04) &  8.3~(10.4$\times$) & / & >3h & / & >3h & / & >3h \\
                $\Hyper_\kw{noEPG}$ & GPU & Blocker & 0.94~(+0) & 9.9~(12.4$\times$) & / & $>$3h & 0.80~(+0) & 1904.1~(142$\times$) & 0.92~(+0) & 2408.6~(523.6$\times$) \\
                $\Hyper_\kw{noHO}$ & GPU & Blocker & 0.94~(+0) & 9.5~(11.9$\times$) & 0.96~(+0) & 472.6~(13.2$\times$) & 0.80~(+0) & 45.0~(3.4$\times$) & 0.92~(+0) & 35.9~(7.8$\times$) \\
                \rowcolor{yellow} \Hyper & GPU & Blocker & 0.94 & 0.8 & 0.96 & 35.7 & 0.80 & 13.4 & 0.92 & 4.6 \\
                \midrule
                \kw{Dedoop} & CPU & Blocker+Matcher & 0.90~(-0.08) & 59.4~(9.4$\times$) & 0.67~(-0.29) & 534.0~(15.0$\times$) & 0.80~(-0.08) & 7643.4~(6.5$\times$) & / & $>$3h \\
                \kw{DisDedup} & CPU & Blocker+Matcher & 0.45~(-0.53) & 94.0~(14.9$\times$) & 0.67~(-0.29) & 644.0~(18.0$\times$) & 0.06~(-0.82) & 917.0~(0.8$\times$) & / & $>$3h \\                
                $\kw{Ditto}_\kw{top2}$ & GPU & Blocker+Matcher &  0.98~(+0) & 9.0~(1.4$\times$) & 0.79~(-0.17) & 6741.2~(188.8$\times$) & 0.88~(+0) & 2308.6~(2.0$\times$) & 0.97~(+0.03) & 381.8~(2.1$\times$) \\
                $\kw{DeepBlocker}_\kw{Ditto}$ & GPU & Blocker+Matcher & 0.99~(+0.01) & 12.4~(2.0$\times$) & / & >3h & / & >3h & / & >3h \\
                \rowcolor{yellow} $\Hyper_\kw{Ditto}$ & GPU & Blocker+Matcher &  0.98 &  6.3 &  *0.96 & *35.7 & 0.88 & 1179.0 & 0.94 & 180.6 \\
                
                \bottomrule
			\end{tabular}
			\par}
	\end{minipage}	
        \vspace{-2ex}
\end{table*}

\vspace{-0.5ex}
\etitle{Measurements}\label{subsec:matrics}.
Following typical ER settings,
\revise{we measured the performance of each method  (blocker, matcher, or the combination of the two) in terms of the runtime and the F1-score, defined as  F1-score = $\frac{2\times\kw{Prec}\times\kw{Rec}}{\kw{Prec} + \kw{Rec}}$.
Here 
$\kw{Prec}$ is the ratio of correctly identified tuple pairs to all identified
pairs and $\kw{Rec}$ is the ratio of correctly identified
tuple pairs to all pairs that refer to the same real-world entity. 
All methods aim to achieve high $\kw{Rec}$, $\kw{Prec}$ and F1-scores.
Following~\cite{DeepBlocking}, we also report the candidate set size ratio~(CSSR), defined as $\frac{|\kw{Ca}(P)|}{|P|\times |P|}$,
\reviseY{when comparing \Hyper with  \kw{DeepBlocker}}, 
to show the portion of tuple pairs that require further comparison by the matcher, 
\ie
the smaller the CSSR, the better the blocker.
} \looseness=-1


\vspace{-0.5ex}
\etitle{Environment.}\label{subsec:env} We run experiments on a Ubuntu 20.04.1 LTS machine powered with 2 Intel Xeon
Gold 6148 CPU @ 2.40GHz, 4TB Intel P4600 PCIe NVMe
SSD, 128GB memory, and 8 Nvidia Tesla V100 GPUs with \reviseX{the widely adopted hybrid cube-mesh topology (see more in~\cite{v100})}.
\eat{32 GB of memory} The programs were compiled with
CUDA-11.0 and GCC 7.3.0 with -O3 compiler. 
\kw{DisDedup}, \kw{SparkER}, and \kw{Dedoop} were run on a cluster of 30 HPC servers, powered with 2.40GHz Intel Xeon Gold CPU, 4TB Intel P4600 SSD, 128GB memory.
\looseness=-1

\vspace{-0.5ex}
\etitle{Default parameters.}
Unless stated explicitly, \CR{we used the following parameters,
best-tuned on each dataset
\reviseS{via \emph{gird search}~\cite{jimenez2008finding}.}
\eat{Intuitively, gird search is a practical method for
systematically exploring the parameter space to get the set of parameters that results in the best performance.}}
\revise{The maximum number of predicates in an \MD is 10.
The number $m$ of data partitions 
is given in Table~\ref{tab:dataset}.}
The sizes of intervals and sliding windows, namely $n_t$ and $n_w$, are 256 and 1024, respectively.
We adopted a regression model as $\N$, with 3 hidden layers,
with 2, 6, and 1 neurons, respectively. 
We used ReLU~\cite{relu} as the
activation function and Adam~\cite{adam} as the optimizer.
We used one GPU by default.
\looseness=-1

\vspace{-0.4ex}
\stitle{Experimental results.}
\reviseX{For lack of space, we report our findings
 on some datasets as follows;
consistent on other datasets\eat{ 
(more in~\cite{full})}.} \looseness=-1

\vspace{-0.4ex}
\stitle{Exp-1: Motivation study.}
We motivate our study by comparing  \Hyper, our rule-based blocker, with the SOTA DL-based blocker \kw{DeepBlocker} (Table~\ref{table:dl_vs_rule}),
where the bracket next to a metric of \kw{DeepBlocker} gives its difference or \reviseC{deterioration} factor \reviseY{to ours.}
\looseness=-1

\vspace{-0.4ex}
\etitle{DL-based blocking vs. rule-based blocking.}
We report recall, CSSR, runtime, and (host and \reviseX{device}) memory for both methods. 
Consistent with \cite{DeepBlocking}, for \kw{DeepBlocker}, each tuple was paired with top-$K$ similar tuples 
as initial candidate pairs,
\reviseY{where $K = 5$ on all datasets (except \kw{DBLP}-\kw{Scholar} where $K=150$).}
As remarked in Section~\ref{sec-intro},
both methods have strengths.
(1) \Hyper effectively reduces the number of pairs to further compare while maintaining high \kw{Rec} (>93\%), \eg
its average \kw{CSSR} is 5.8\textpertenthousand~less than \kw{DeepBlocker}.
(2) \Hyper is at least 10$\times$ faster. 
(3) \Hyper consumes less memory than \kw{DeepBlocker},
\eg the host memory it consumes is at least 23.3$\times$ less than  \kw{DeepBlocker}. 
(4) Note that the \kw{Rec} of \Hyper is slightly lower than \kw{DeepBlocker},
which is acceptable given its convincing speedup and memory saving,
since the primary goal of a blocker is to improve the efficiency and scalability of ER,
not to improve the accuracy of ER (the goal of a matcher). \looseness=-1

\vspace{-0.4ex}
\stitle{Exp-2: Accuracy-efficiency.}
\revise{We report the F1-scores and runtime of all blockers and integrated ER solutions (\ie blocker + matcher) in Table~\ref{tab:overview}.
Here
\kw{DeepBlocker} pairs each tuple with its top-2 tuples as initial candidate pairs.
For all blockers,
the bracket next to each F1-score (resp. time) gives the difference (resp. \reviseC{slowdown}) in  F1-score (resp. time) to \Hyper (marked yellow).
For a fair comparison, 
the brackets of each integrated ER solution give the \reviseC{difference} compared with $\Hyper_\kw{Ditto}$ (marked yellow).} \looseness=-1
 \looseness-1

\etitle{Accuracy.}
\revise{We mainly analyze the F1-scores of \Hyper, which are consistently above 0.8 over all datasets.}
Besides, we find:

\sstab
(1)  \Hyper outperforms CPU-based distributed solutions,
\eg it achieves \revise{up to 0.29, 0.74, and 0.72} improvement in F1-score against \kw{Dedoop}, \kw{DisDedup}, and \kw{SparkER}, respectively,
\revise{even though the former two are integrated with matchers.}
This is because these solutions exploit data partition-based parallelism only,
\reviseC{which may lead to false negatives
if matched tuples are put into different partitions.}
\looseness=-1


\vspace{-0.2ex}
\sstab
(2) Compared with the four GPU-based baselines,
\Hyper has comparable accuracy.
In particular, it even beats $\kw{Ditto}_\kw{top2}$,
the SOTA matcher, by \revise{0.17} F1-score in \kw{IMDB}.
This shows that \revise{even without a matcher}, \Hyper alone is already accurate in certain cases. 
Moreover, \kw{DeepBlocker} and $\kw{DeepBlocker}_\kw{Ditto}$ struggle to handle large datasets. \reviseY{When facing million-scale data, they 
cannot finish in 3 hours.} 
This again motivates the need for rule-based alternatives.

\vspace{-0.2ex}
\sstab
(3) Combing \Hyper with $\kw{Ditto}$, 
$\Hyper_\kw{Ditto}$ further boosts the accuracy,
achieving the best F1-score  in \kw{Songs}.
\revise{Nevertheless, DL-based solutions still have the best F1 scores in other cases,
\reviseY{justifying that 
none of them} can dominate the other in all cases.} \looseness=-1

\sstab
(4) $\Hyper_\kw{noEPG}$ and $\Hyper_\kw{noHO}$ are as accurate as \Hyper,
since they only differ in the optimizations.

\eat{ 
The table shows that 
(1) all variants  \Hyper can achieve an high \kw{Prec} with relative small candidate set~(low \kw{recall}), 
its F1-score over benchmark datasets is consistently above 0.8.
Surprisingly, for \kw{IMDB}, 
\Hyper beats all baselines in term of accuracy, \eg it outperform $\kw{Ditto}_\kw{top2}$ by 17\%,  even without the  subsequently matching step.
This justifies the need for rule-based blocking;
(2) By combing blocking in \Hyper with  $\kw{Ditto}_\kw{top2}$, $\Hyper_\kw{Ditto}$ maintain comparable accuracy.
It achieves the best F1-score on \kw{DBLP}-\kw{ACM}, \kw{IMDB}, \kw{Songs}.
(3) \Hyper achieves 9.8\%, 55.3\%, and 56.5\% improvement for \kw{Dedoop}, \kw{DisDedup}, and \kw{SparkER} on average.
The main reason is that distributed solutions exploit partition parallelism only.
With the increasing number of data partitions, it would result in mismatches.
}

\vspace{-0.2ex}
\etitle{Runtime.}
We next report the runtime. 
(1) \Hyper runs substantially faster than all baselines, 
\eat{since it \reviseY{generates effective execution plan} and explores novel optimization and scheduling strategies under the shared memory architecture,}
\eg
\reviseY{
it is at least 6.8$\times$, 9.1$\times$, 10.4$\times$, 15.0$\times$, 18.0$\times$, 11.3$\times$ and 15.5$\times$ faster than \kw{SparkER}, \kw{GPUDet}, \kw{DeepBlocker}, \kw{Dedoop}, \kw{DisDedup}, \kw{Ditto}, and $\kw{DeepBlocker}_\kw{Ditto}$  respectively}.
(2) $\Hyper_\kw{Ditto}$ is slower than \Hyper as expected since it performs additional matching.
Nonetheless,
$\Hyper_\kw{Ditto}$  is at least 1.4$\times$ (resp. 2.0$\times$) faster than $\kw{Ditto}_\kw{top2}$ (resp. $\kw{DeepBlocker}_\kw{Ditto}$). 
Given its comparable F1-score,  
we substantiate our claim (Section~\ref{sec-intro}) that blocking is a crucial part of the overall ER process.
(3) \Hyper is at least \revise{12.4$\times$ and 3.4$\times$} faster than $\Hyper_\kw{noEPG}$ and $\Hyper_\kw{noHO}$, respectively,
verifying the usefulness of execution plans and hardware optimizations. 
\looseness=-1

\begin{figure*}[t]
    \subcaptionbox{
		\revise{\kw{NVC}: Impact of $m$}
	}{
		\includegraphics[width=0.24\textwidth]{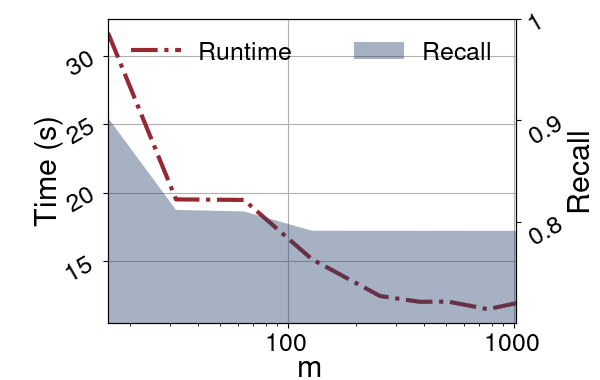}\vspace{-1.5ex}
	}	    	    
    \hspace{-3mm}
	\subcaptionbox{
		\revise{\kw{TFACC}-\kw{Large}: Varying \#GPUs}
	}{
        \includegraphics[width=0.24\textwidth]{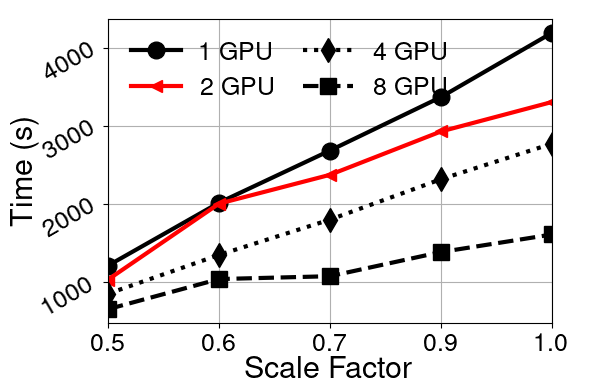}\vspace{-1.5ex}
	}
 	\hspace{-3mm}
        \subcaptionbox{
		\revise{\kw{IMDB}: Impact of schedulers}
	}{
		\includegraphics[width=0.24\textwidth]{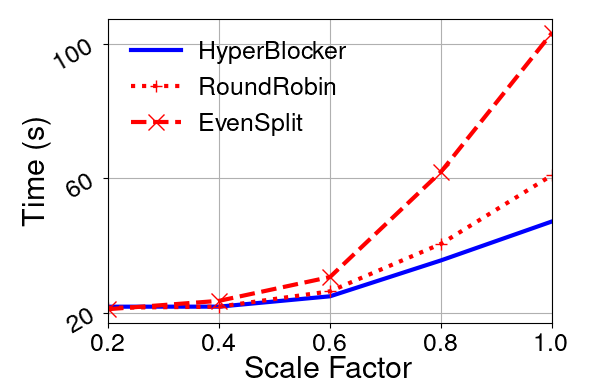}\vspace{-1.5ex}
	}
 	\hspace{-3mm}
    \subcaptionbox{
		\kw{TFACC}: Varying $|\varphi|$
	}{
		\includegraphics[width=0.24\textwidth]{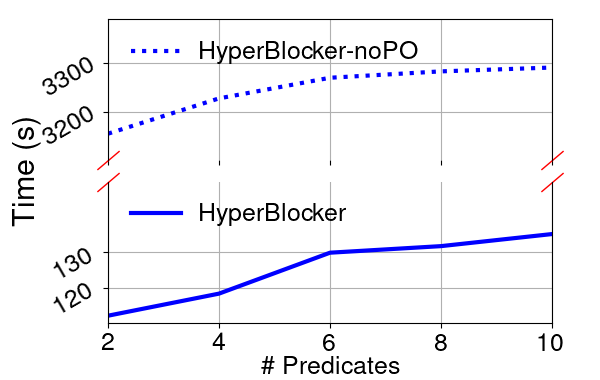} \vspace{-1.5ex}
	}
        \subcaptionbox{
		\revise{\kw{TFACC}: Varying $|\Delta|$}
	}{
		\includegraphics[width=0.24\textwidth]{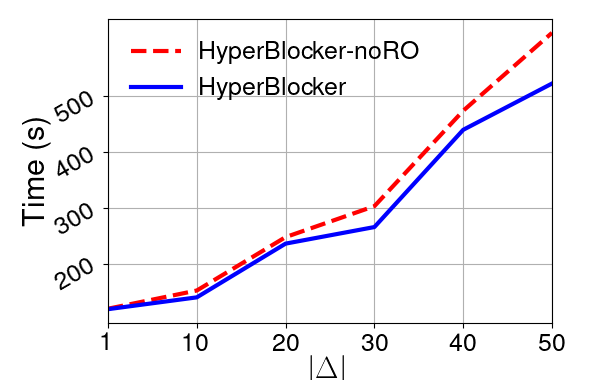} \vspace{-1.5ex}
	}
	\hspace{-3mm} 
	\subcaptionbox{
		\reviseX{\kw{DBLP}-\kw{ACM}: Varying noise\%}
	}{
		\includegraphics[width=0.24\textwidth]{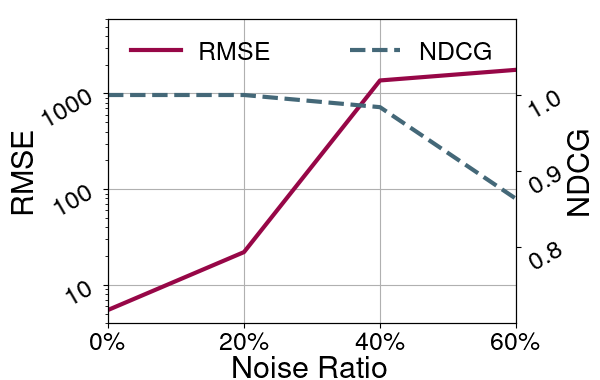}\vspace{-1.5ex}
	}
        \hspace{-3mm} 
        \subcaptionbox{
		\reviseX{More ordering strategies}
	}{
		\includegraphics[width=0.24\textwidth]{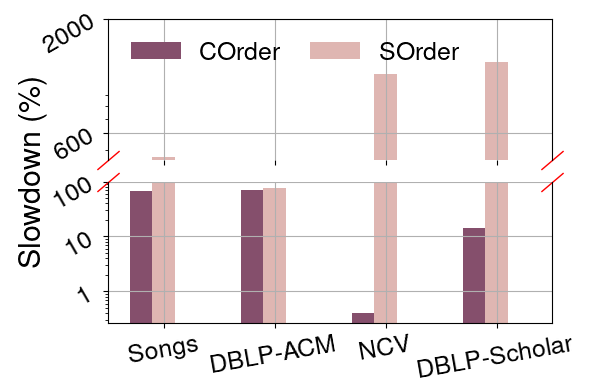}\vspace{-1.5ex}
	}
	\hspace{-3mm} 
        \subcaptionbox{
		Ablation study
	}{
		\includegraphics[width=0.24\textwidth]{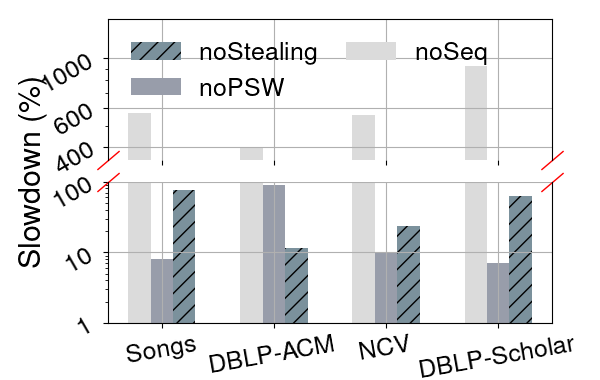}\vspace{-1.5ex}
	}
    \vspace{-3ex}
    \caption{Efficiency, scalability, and effectiveness of \Hyper} \label{exp:scalablity}
    \vspace{-3ex}
\end{figure*}

\vspace{-0.2ex}
\etitle{Impact of $m$.}
\reviseY{Figure~\ref{exp:scalablity}~(a) reports how the number $m$ of data partitions affects the recall (the right y-axis)  and the runtime  (the left y-axis) on \kw{NVC}.}
As shown there, both metrics of \Hyper
decreases with increasing $m$.
This is because when there are more partitions,
both the number of pairwise comparisons and the candidate matches that can be identified in each partition are reduced.

\eat{
\begin{figure*}[t]
        \subcaptionbox{
		\kw{TFACC}: Varying $|\varphi|$
	}{
		\includegraphics[width=0.24\textwidth]{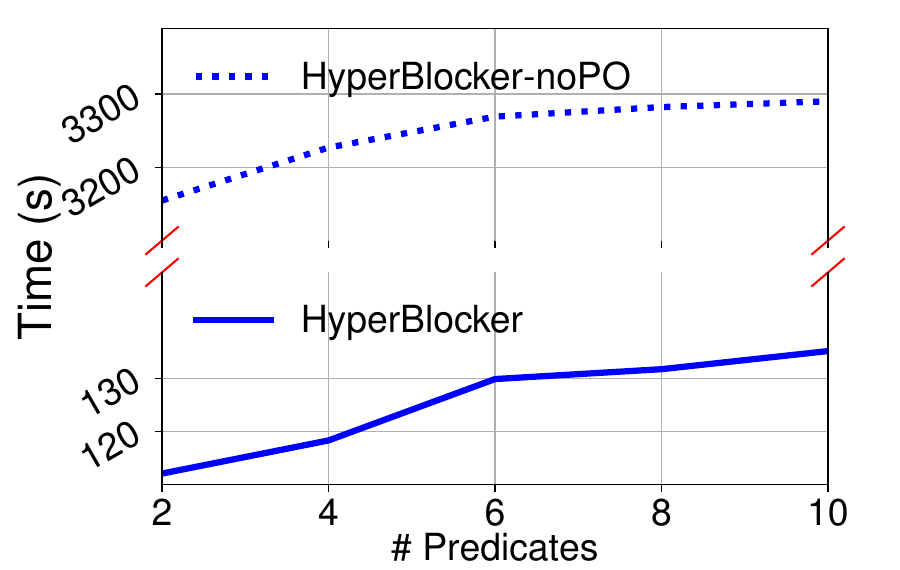} \vspace{-1.5ex}
	}
	\hspace{-3mm} 
	\subcaptionbox{
		\revise{\kw{TFACC}: Varying $|\Delta|$}
	}{
		\includegraphics[width=0.24\textwidth]{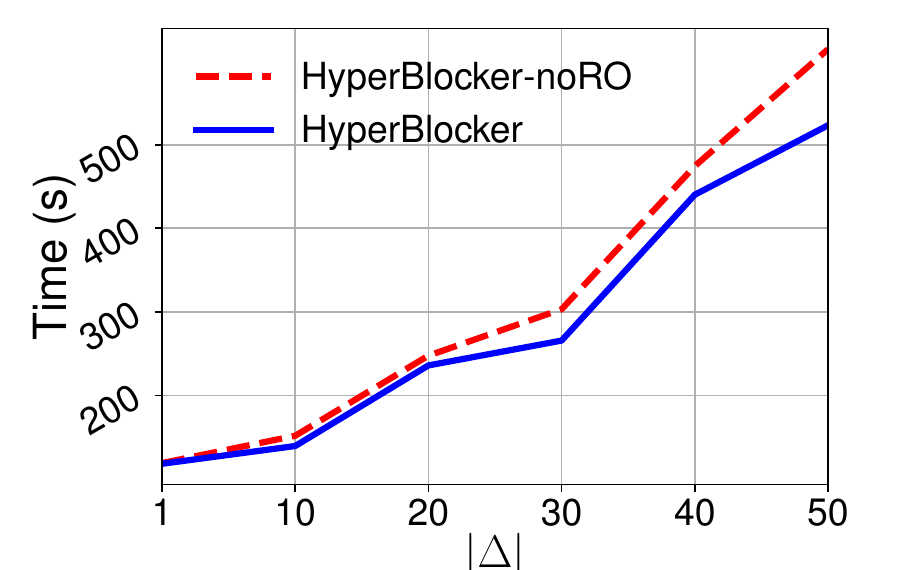} \vspace{-1.5ex}
	}
        \hspace{-3mm} 
        \subcaptionbox{
		\reviseX{\kw{DBLP}-\kw{ACM}: Varying noise\%}
	}{
		\includegraphics[width=0.24\textwidth]{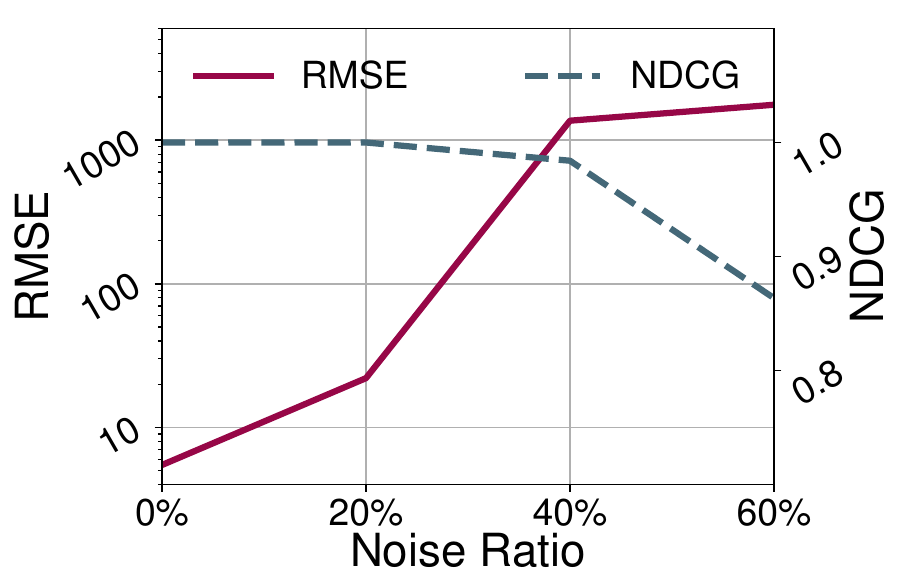}\vspace{-1.5ex}
	}
	\hspace{-3mm} 
        \subcaptionbox{
		\reviseX{More ordering strategies}
	}{
		\includegraphics[width=0.24\textwidth]{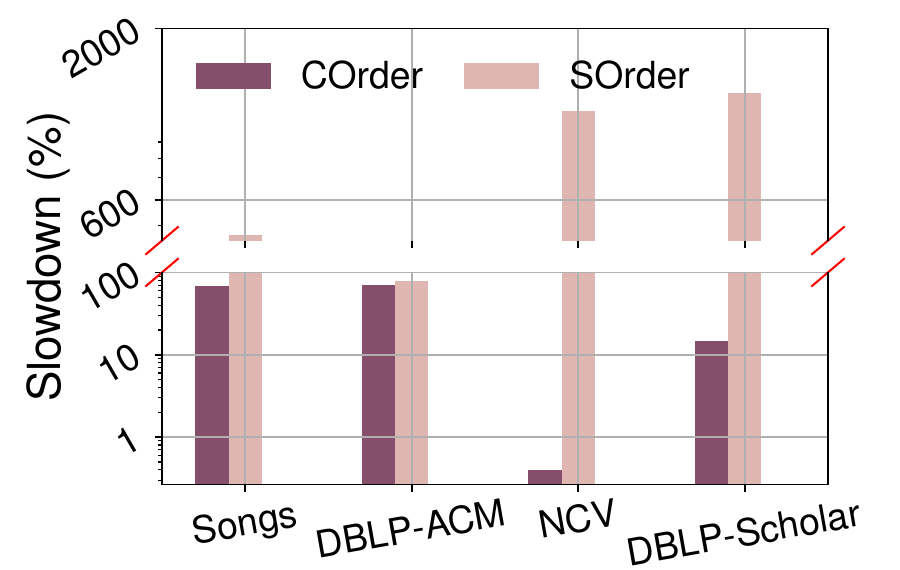}\vspace{-1.5ex}
	}
    \vspace{-3ex}
    \caption{\warn{Experiments on EPG}} \label{exp:EPG}
    \vspace{-2ex}
\end{figure*}
}

\vspace{-0.4ex}
\stitle{Exp-3: Scalability.}
\reviseX{We tested our scalability under multi-GPUs scenarios. 
The default number of GPUs is 4 in this set of experiments.}
\looseness=-1

\vspace{-0.2ex}
\etitle{Varying $|D|$/\#GPUs.}
We varied the scale factor of $D$ in \revise{\kw{TFACC_{large}}}
and tested \Hyper with different numbers of GPUs in Figure~\ref{exp:scalablity}(b).
\Hyper scales well with data sizes,
\eg
with 8 GPUs, it takes 1604s to process \revise{36M} tuples;
\reviseX{this is not feasible for both CPU- and GPU-based baselines.}
\eat{\Hyper can complete a blocking task on a table with 36 million tuples within half an hour, a performance that neither existing CPU-based nor DL-based competitors can achieve.}
When the number of GPUs changes from 1 to 8, 
\Hyper is 2.6$\times$ faster,
\reviseX{
since 
\Hyper mainly accelerates the operations on GPUs, 
while other parts of the system (\eg I/O and data partitioning) may also limit the overall performance.}

\vspace{-0.2ex}
\etitle{Impact of task schedulers.}
We tested the impact of task schedulers,
by comparing \Hyper with two variants,
that uses \kw{EvenSplit} and \kw{RoundRobin} for scheduling (Section~\ref{subsec:GPUs}), respectively,
\reviseC{by varying $|D|$
 in Figure~\ref{exp:scalablity}(c).}
\Hyper works better than the two, \eg
when the scale factor is 100\%, \Hyper is 1.3$\times$ and 2.2$\times$ faster than \kw{RoundRobin} and \kw{EvenSplit}, respectively,
\reviseX{since both variants may limit CUDA’s ability to dynamically schedule tasks.}

\vspace{-0.4ex}
\reviseX{ 
\stitle{Exp-4: Tests on EPG (Section~\ref{sec-plan}).}
We evaluated EPG (and its offline model $\N$) and justified the need of effective evaluation orders.} 

\vspace{-0.2ex}
\etitle{Varying $|\varphi|$.}
We tested the number $|\varphi|$ of predicates in each \MD $\varphi$
\reviseY{against $\Hyper_\kw{noPO}$ that evaluates predicates in a random order}
in \kw{TFACC} (Figure~\ref{exp:scalablity}(d)).
(1) \Hyper takes longer with larger $|\varphi|$, as expected.
(2) \Hyper is feasible in practice,
\eg when $|\varphi|=10$, it only takes 135.2s. 
(3) On average, \Hyper shows 32.5$\times$ speedup to $\Hyper_\kw{noPO}$.
This justifies the importance of predicate ordering in efficient rule-based blocking.

\vspace{-0.2ex}
\etitle{Varying $|\Delta|$.}
We evaluated the impact of the number $|\Delta|$ of \MDs in $\Delta$ in Figure~\ref{exp:scalablity}(e), where
\Hyper takes longer with more rules, \eg
it takes  \revise{523.2}s when $|\Delta|=$  \revise{50},
\reviseY{and consistently beats $\Hyper_\kw{noRO}$,
a variant  that evaluates rules in a random order.}\looseness=-1

\reviseX{ 
\vspace{-0.2ex}
\etitle{Shallow model $\N$.}
We evaluated the performance of $\N$ in EPG by (1) its sensitivity to noises, (1) the resulting predicate ordering,
compared with the ``ground truth'' ordering derived from  actual costs, and
(3) the speedup of estimating the actual costs using $\N$.

\vspace{-0.3ex}
\sstab
(1) Given a noise ratio $\beta\%$, we injected $\beta\%$ noises to training data of $\N$, to disturb its distribution,
and report RMSE (Root Mean Squared Error),
a widely used metric for regression, in Figure~\ref{exp:scalablity}(f) (the left y-axis).
The RMSE of $\mathcal{N}$ does not degrade much
when $\beta\%$ = 20\%.
However,
when $\beta\%$ continues to increase, $\mathcal{N}$ becomes inaccurate. 
\eat{A case study about resulting orderings under different $\beta\%$ is given in~\cite{full}.}
\looseness=-1

\vspace{-0.3ex}
\sstab
(2) We compared the predicate ordering estimated via $\N$ with the ground truth one using NDCG (Normalized Discounted Cumulative Gain~\cite{NDCG}), a widely used metric for evaluating ranking, in Figure~\ref{exp:scalablity}(f) (the right y-axis).
The result shows that the two orderings are close (\ie NDCG is high), even when the noise ratio is 40\%. 
\looseness=-1

\vspace{-0.3ex}
\sstab
(3) The average time for computing the actual cost of a predicate is 0.8s on
\kw{DBLP}-\kw{ACM}, as opposed to 0.007s \CR{for the estimated cost}\eat{  (see more in~\cite{full})}.
}
\looseness=-1

{
\vspace{-0.3ex}
\etitle{More ordering strategies.}
To justify the need for both cost and effectiveness,
we compared two more strategies using designated \MDs:
(1) $\kw{COrder}$,  that prioritizes cheap predicates
(\eg always evaluate equality first, 
\reviseS{a common strategy in existing DBMS, as remarked in Section~\ref{sec-plan}}) and 
(2) $\kw{SOrder}$, that prioritizes selective predicates.
\reviseS{For all orders, we applied the same partitioning strategy (Section~\ref{subsec:GPUs}).}
To better visualize the effects on different datasets,  
we report the slowdown percentages 
in Figure~\ref{exp:scalablity}(g).
$\kw{SOrder}$~(resp. $\kw{COrder}$) is on average slowed by 733.6\% 
(resp. 38.2\%)
compared with \Hyper.
This said, we strike a balance between the two strategies.}
\looseness=-1

\eat{This is because: 
(a) the main goal of cardinality estimation 
is to  minimize intermediate results, thus 
$\kw{CEOrder}$ is easily making predicates with high discrimination power performing early. However, these predicates often come with higher computational complexity.
(b) $\kw{SDOrder}$ prioritizes predicates based solely on their computational cost, leading to the early execution of predicates with high speed but low discrimination power resulting in redundant evaluations. 
In contrast, \Hyper learns from the data and considers both factors to determine the most "suitable" orderings.
}

\reviseX{ 
\vspace{-0.5ex}
\stitle{Exp-5: Tests on hardware optimizations (Section~\ref{sec-exec-model}).}
Finally,
we conducted an ablation study on hardware optimizations 
and report the runtime statistics.
We compared three baselines:
(1)  \kw{noSeq}, that recursively implements DFS without sequential execution paths,
(2) $\kw{noPSW}$ that assigns continuous intervals to each TB without parallel sliding windows, and
(3) $\kw{noStealing}$, where GPUs automatically schedule a new TB whenever one is done,
without task stealing.
To better visualize the effect, 
below we used $D$ as a single partition.}
\looseness=-1

\reviseX{

\vspace{-0.3ex}
\etitle{Ablation study.}
We show the slowdown percentages compared to \Hyper in Figure~\ref{exp:scalablity}(h).
We find:
(1) \kw{noSeq} is much slower than \Hyper, since recursive DFS is not efficient on GPUs.
(2) $\kw{noStealing}$ and $\kw{noPSW}$ are on average 
\reviseX{43.1\%} and \reviseX{28.8\%}
slower than \Hyper, respectively, 
justifying the use of both optimizations.


\reviseX{
\vspace{-0.3ex}
\etitle{Runtime statistic.}
We adopted NSight~\cite{Nsight}, a profiling tool provided by NVIDIA, 
and report
\emph{wait stalls} (\ie the number of clock cycles that the kernel spent on waiting),
\emph{branch efficiency} (\ie  the ratio of correctly predicted branch instructions), and 
the \emph{average number active threads per warp} in \kw{TFACC} (Table~\ref{tab:statistic}). 
\eat{In particular:
\mbi 
\item \emph{Wait stalls} refer to periods where the execution of instructions is delayed due to various dependencies.
\item \emph{Branch efficiency} states the ratio of uniform control flow decisions over all executed branch instructions. Higher values are better, as warps more often take a uniform code path. 
\item \emph{Avg. active threads per warp} indicates what fraction of threads of warps in a kernel have been active on average. It can also be seen as a measure that is inversely proportional to the overall thread divergence.
\mei}
\Hyper performs the best in all metrics. 
\eat{The reasons are twofold: (1) when evaluating repeated predicates within a \emph{sequential execution path}, potential branch divergence is avoided, and the number of operations is reduced; (2) a \emph{sequential execution path} benefits from better data locality and avoids stack/backtracking for each thread, resulting in fewer stalls. }
The reasons are twofold:
(1) while divergence is sometimes unavoidable, a recursive DFS exacerbates it (\eg due to stacking), leading to more idle threads; 
and
(2) the workloads can be imbalanced,
\eg without parallel sliding windows,
\kw{noPSW} incurs a larger number of wait stalls compared with \Hyper.


}

}

\vspace{-0.6ex}
\stitle{Summary.}
We find the following.
(1) \Hyper outperforms \revise{prior blockers and integrated ER solutions.} 
\reviseY{
It is at least 6.8$\times$, 9.1$\times$, 10.4$\times$, 15.0$\times$, 18.0$\times$, 11.3$\times$ and 15.5$\times$ faster than \kw{SparkER}, \kw{GPUDet}, \kw{DeepBlocker}, \kw{Dedoop}, \kw{DisDedup}, \kw{Ditto}, and $\kw{DeepBlocker}_\kw{Ditto}$  respectively}.
(2) By combining \Hyper with \kw{Ditto}, 
we save at least 30\% of time with comparable accuracy.
(3) \Hyper beats all its variants (except $\Hyper_\kw{Ditto}$) in both runtime and accuracy,
justifying the usefulness of various optimizations:
(a) \reviseY{EPG} specifies an effective evaluation order, improving the runtime by at least 12.4$\times$ and 
(b) the hardware optimizations on GPUs speedup blocking by at least 3.4$\times$.
(4) \Hyper scales well with various parameters, \reviseY{\eg 
it completes blocking in 1604s on 36M tuples.}
\looseness=-1

\begin{table}[t]
\caption{\reviseX{Runtime info ( $\uparrow$: higher is better vs. $\downarrow$: lower is better)}}\label{tab:statistic}
\vspace{-2ex}
\setlength{\tabcolsep}{5pt}
\centering
\scriptsize
\begin{tabular}{cccc}
\toprule
      \textbf{Method} & \textbf{\tabincell{c}{$\downarrow$ Wait stalls (in \\ terms of clock cycles)}} & \textbf{\tabincell{c}{$\uparrow$ Branch \\ efficiency}} & \textbf{\tabincell{c}{ $\uparrow$ Average number of \\ active  threads per warp}} \\
\midrule
\kw{noSeq}  & 4.25  &  89.9\%  &   14.45 \\
\kw{noPSW}  & 13.79  &  96.3\%   &  25.59 \\
\kw{noStealing}  & 4.11  &  96.2\%   &  27.62 \\
\Hyper & 4.07 &  96.4\%   &   28.21  \\        \bottomrule                         
\end{tabular}
\vspace{-5ex}
\end{table}
\section{Related Work} \label{sec:related-work}

We categorize the related work in the literature as follows.

\vspace{-1ex}
\stitle{Blocking algorithms.}
There has been a host of work on the blocking algorithms, classified as follows:
(1) Rule-based~\cite{dedoop,dis-dup,rules-blocking,papadakis2011compare,gu2004adaptive},
\eg 
\cite{gu2004adaptive} 
creates data partitions 
and then refines candidate pairs in every partition, 
by removing mismatches with similarity measures or length/count filtering \cite{mann2014pel}.
(2) DL-based ~\cite{DeepER,DeepBlocking,zhang2020autoblock,javdani2019deepblock},
which cast the generation of candidate matches into a binary classification problem, where each tuple pair is labeled ``likely match'' or ``unlikely match'',
\eg 
\cite{DeepBlocking} adopts similarity search to
generate candidate matches for each tuple based on its top-$K$ probable matches in an 
embedding space.
\revise{DL-based blocking and rule-based blocking share the same goal, but are different in their approaches, where the former 
focuses on learning the distributed representations of tuples,
while the latter emphasizes explicit logical reasoning. 
}

\reviseY{Although we study rule-based blocking,
we are not to develop another blocking algorithm.
Instead,
we provide} a GPU-accelerated blocking solution.
As a testbed, we use \MDs as our blocking rules,
which subsume many existing rules~\cite{konda2016magellan,papadakis2011compare} as special cases.

\vspace{-0.8ex}
\stitle{Parallel blocking solvers.}
Several parallel blocking systems have been proposed, \eg \cite{dedoop,dis-dup, MinoanER,sparkER,kolb2012load,collective-large,MPC-ER,preedet,parallelProgressiveER,Falcon},
mostly under MapReduce \cite{dedoop, dis-dup, sparkER} or MPC \cite{MPC-ER,preedet,deng2022deep},  which aim at scaling to large data with a cluster of machines.
\kw{DisDedup}~\cite{dis-dup} uses a triangle distribution strategy to minimize both comparisons and communication over Spark\cite{spark}.
\kw{Minoan}~\cite{MinoanER} runs on top of Spark and applies parallel meta blocking~\cite{efthymiou2015parallel} to minimize its overall runtime.
\looseness=-1

This work differs as follows. 
Unlike MapReduce-based systems, which split data at the \reviseX{coordinator}  and execute tasks on workers, \Hyper focuses on collaborating GPUs and CPUs, to promote better resource utilization and massive parallelism. 
\reviseY{\Hyper is designed for the shared memory architecture of GPUs and is fine-tuned to exploit GPU hardware for rule-based blocking. To the best of our knowledge, incorporating both GPU and CPU characteristics has not been considered in prior parallel blocking solutions.}
\looseness=-1

\vspace{-0.8ex}
\stitle{GPU-accelerated techniques.}
GPUs have been used extensively 
to speed up the training of DL tasks.
Recent works exploit GPUs to accelerate data processing, \eg GPU-based query answering \cite{Hardware-conscious-Hash-Joins-on-GPUs,gpu-query,gpu-query2} and similarity join~\cite{nn-gpu,sim_join_gpu,Faiss}.  
Closer to this work are~\cite{sim_join_gpu,Faiss} which leverage GPUs for similarity join,
since blocking can be regarded as a similarity join problem under the assumption that two tuples refer to the same entity if their similarity is high. 
Similarity join is often served as a preprocessing step of ER.

In contrast, \Hyper aims at expediting rule-based blocking, addressing challenges in rule-based optimization that are not incurred in similarity join.
The closest work is \kw{GPUDet}~\cite{GPU-ER}, which employs GPUs to expedite similarity measures. 
\Hyper differs from  \kw{GPUDet}, 
in its data/rule-aware execution plan designated for rule evaluation, beyond similarity measures. 
It also incorporates hardware-aware optimizations for improving GPU utilization.
\looseness=-1

 \eat{
Also, closer to this work is \cite{GPU-ER}.  \cite{GPU-ER} proposes a GPU-based duplication detection algorithm that calculates edit-based similarity on GPU. 
However, \cite{GPU-ER} can not scale to large datasets because it has to perform multiple non-overlapping comparison rounds under the Fermi framework~\cite{}. 
Moreover, we show that switching edit-based similarity computation to the GPU without hardware-level optimizations leads to low utilization of GPU.
In contrast, \Hyper
(1) proposes an asynchronous coprocessing pipelined architecture to “cancel” I/O costs caused by slow PCIe
(2) employs a three-level parallelism strategy in \Hyper to  maximize GPU efficiency;
(3) develops new GPU hardware-conscious optimization strategies including matching order advisor, checkpoints, and scheduling~(see Section \tbf for more details).
}

\vspace{-0.6ex}
\stitle{Query optimizations.}
\reviseS{
Also related  to EPG  is query optimization in DBMS~\cite{neo,simjoin_order,rheinlander2017optimization,MQO,larson2007cardinality,moerkotte2006building}, which uses sampling, statistics, or profiling to get execution plans via cost and cardinality estimation. 
Since rule-based blocking is in DNF, with arbitrary similarity comparisons and multiple rules,
EPG is particularly related to the optimizations on DNF SQLs with UDFs~\cite{udf1,UDF_tutorial,tuplex,simjoin_order},
\eg~\cite{tuplex} analyzes Python UDFs to reorder operators based on data/operation types.
\looseness=-1

\eat{Unfortunately,
many optimizers in DBMS struggle when dealing with UDFs and {\tt OR} operations.
For example, SQL Server~\cite{sql_server} restricts UDFs to a single thread, and PostgreSQL~\cite{postgresql} treats UDFs as black boxes since it is hard to accurately estimate their runtime performance, which may depend on specific functions, thresholds, and data distributions. 
The most relevant related work is TupleX~\cite{tuplex}, which analyzes Python UDFs to reorder operators. However, it is limited to type analysis (e.g., None, string, or integer).} 
\looseness=-1



\eat{
estimating the cost of UDFs rely on 

due to the difficulty of 

rely on accurate statistics and cost models, but with UDFs, estimating selectivity is challenging due to their dependence on specific functions, thresholds, and data distributions.

\eg 
}


EPG differs from existing query optimizations:
(1) EPG optimizes the execution, no matter what comparisons (\eg equality or similarity) are adopted,
while many DBMS optimizers struggle when similarity comparisons are encoded as UDFs, \eg SQL Server~\cite{sql_server} restricts UDFs to a single thread, and PostgreSQL~\cite{postgresql} treats UDFs as black boxes. 
This said, EPG solves a more specialized
 problem, beyond general query optimization, for arbitrary comparisons.
(2) It is hard for most optimizers to accurately estimate the runtime performance of UDFs~\cite{rheinlander2017optimization}, \CR{which may depend on specific measures/data,
while we consider the time/selectivity of predicates, using learned and LSH-based models for accurate estimation.} 
(3) EPG employs tree structures and bitmaps, to effectively handle the disjunction logic behind blocking and to reuse computation,
while traditional DBMS may be forced to perform full scan when evaluating {\tt OR} operations.
(4) EPG produces a data partitioning scheme based on the execution tree as a by-product, to coordinate  across multiple GPUs.
\eat{More in-depth analysis about  EPG with the plans generated with DMBS and UDFs optimizations can be found in~\cite{full}.}
\looseness=-1

}

\section{Conclusion}
\label{sec-conclude}
The novelty of \Hyper consists of (1) a pipelined architecture that overlaps the data transfer from/to CPUs
and the operations on GPUs;
(2) a data-aware and rule-aware execution plan generator on CPUs, that
specifies how rules are evaluated;
(3) a variety of hardware-aware optimization strategies that achieve massive parallelism, by exploiting GPU characteristics; and
(4) \reviseY{partitioning and scheduling} strategies to achieve workload balancing across multiple GPUs.
Our experimental study has verified that \Hyper is much faster than existing  CPU-powered distributed systems and
GPU-based ER solvers,
while maintaining comparable accuracy.

\eat{We have studied the parallel entity resolution under Heterogeneous System Architecture.  
 We have given \kw{Hyper}, a hybrid CPU-GPU parallel system for entity resolution, 
which  supports a simple programming model  such that users can plug in existing sequential ER algorithms with minor changes.
We have presented a dynamic scheduler that carefully overlaps communication and computation, and scales out on demand.
Our experimental study has shown that our approach is effective. 
One topic for future work is to design a multi-GPU parallel scheme for the collective ER, 
to recursively identifies tuples by making use of matches in the previous rounds.}


There are some future topics:
\eat{(a) accelerate the evaluation of \reviseC{multi-variable rules, \eg\  \REEs~\cite{REE-discovery}}, using GPUs;}
(a) give a different plan on each partition;
\reviseS{(b) explore the materialization of partial evaluation results to avoid divergence and (c)
investigate whether EPG and traditional optimizers can complement/enhance each other.}


	\begin{acks}
		\vspace{-0.4ex}
		We sincerely thank Wenfei Fan, Shuhao Liu,
		Yaoshu Wang, and Weijie Ou for their comments and helpful discussions. 
  
	\end{acks}
 	
	\clearpage
	\balance
	
	\bibliographystyle{ACM-Reference-Format}
	\bibliography{paper}
    
    \clearpage
    \balance

\end{document}